\documentclass[a4paper, 11pt, twoside]{article}
\title{Watching gravitational waves}
\author{Joachim Moortgat}
\date{\today}

\newcommand{\ted}{{\bf e}}
\newcommand{\ric}[2]{\Gamma^{#1}_{\ #2}}
\newcommand{\bi}{\begin{itemize}}
\newcommand{\ei}{\end{itemize}}

\newcommand{\haak}[1]{\left( #1 \right)}
\newcommand{\brak}[1]{\left[ #1 \right]}
\newcommand{\krul}[1]{\left\{ #1 \right\}}
\newcommand{\bq}{\begin{equation}}
\newcommand{\eq}{\end{equation}\noindent}

\newcommand{\half}{\frac{1}{2}}
\newcommand{\kwart}{\frac{1}{4}}
\newcommand{\ddt}[1]{\frac{d #1}{dt}}

\newcommand{\mdoos}[1]{\mbox{\hspace{0.5 cm} #1 \hspace{0.5 cm}}}
\newcommand{\mdoosje}[1]{\mbox{\hspace{0.25 cm} #1 \hspace{0.25 cm}}}
\newcommand{\afg}[1]{\frac{\partial #1}{\partial t}}
\newcommand{\afgg}[1]{\frac{\partial^2 #1}{\partial t^2}}
\newcommand{\afgz}[1]{\frac{\partial #1}{\partial z}}

\newcommand{\btel}{\begin{enumerate}}
\newcommand{\etel}{\end{enumerate}}
\newcommand{\bqa}{\begin{eqnarray}}
\newcommand{\eqa}{\end{eqnarray}\noindent}
\newcommand{\ba}{\begin{array}}
\newcommand{\ea}{\end{array}\noindent}

\usepackage{latexsym}

\usepackage{fancyhdr}
\usepackage{graphicx}
\usepackage{amssymb}
\usepackage[colorlinks]{hyperref}

\begin{document}

\thispagestyle{empty}
\maketitle
\newpage
\thispagestyle{empty}

\begin{abstract}
In this thesis the interaction of gravitational waves ({\sc gw}s) with 
electromagnetic waves ({\sc emw}s) in a static magnetic background field is considered. 
This interaction is a consequence of the general relativistic Einstein field equations. 
These equations dictate the excitation of electromagnetic waves when a
gravitational wave interferes with a static electromagnetic background field and a similar reverse process.

Such interactions can become effective, close to very energetic 
sources such as colliding neutron star binaries (i.e. gamma ray bursts), and quacking supernova 
remnants (magnetars), or in the large scale magnetic fields of the early universe. 
The former two sources have strong, localized and the latter weak, but extended magnetic fields. This is important 
because the energy transfer efficiency of {\sc gw} $\Leftrightarrow $ {\sc emw} conversions appears to be quadratically proportional to the background field strength and the extension of the interaction region.

All calculations in this thesis are done in a general relativistic, space+time, non-coordinate formalism. The 
reason for this is that in such {\em tedrad systems}, equations remain transparent and facilitate easy physical interpretation and connection to measurements.

In this framework, the conversion efficiency of {\sc gw}s to light and vice versa in a vacuum is considered, first in an estimate and then in a more elaborate and general, exact calculation. The {\sc gw}$\Rightarrow ${\sc emw} conversion is proposed as a possible indirect detection device for gamma ray bursts and magnetars. Also, the possibility is considered of explaining the small fluctuations in the  cosmic background radiation by the conversion of {\sc gw}s in the early universe to {\sc emw}s superposed on the homogeneous background radiation.

Next, to obtain more realistic results, the same process is examined in a thin plasma which leads to the same {\sc emw}s as generated in a vacuum. The importance of the presence of this plasma, though, is that it might damp the generated radio waves before they can travel over astromical distances, unless non-linear effects lead to higher frequencies of the {\sc emw}s. If radio waves with large enough frequencies are generated, gamma ray bursts and supernov{\ae} might be detectable with (space based) radio detectors such as the proposed  {\em Astronomical Low Frequency Array} ({\sc alfa}) as well as with {\sc gw} detectors such as {\em Laser Interferometer Gravitational wave Observatory} ({\sc ligo}) with a event rate of as many as a few per year in our local galaxy 
group and the Virgo cluster. 

In the last chapter, an entirely different interaction is proposed, in which the gravitational wave interacts with the plasma and generates fast magneto-acoustic plasma waves, thus dissipating its energy into the plasma. The plasma can then emmit the energy as electromagnetic radiation.  
Theoretical models for gamma ray burst could be improved a lot if even a small fraction of the {\sc gw} energy could be converted into {\sc emw}s in this fashion. The reason for this is that the energy released in a neutron star binary merger is expected 
to be released mainly in {\sc gw}s, whereas the obeserved energy flux is mostly electromagnetic. 
The dispersion relation derived for this interaction is the most interesting, new, result of this thesis work.
\end{abstract}

\newpage
\tableofcontents

\newpage

\section{Introduction and overview}
A consequence of Einstein's General Theory of Relativity is the 
existence of the interesting concept of gravitational waves. These 
waves originate from the most energetic events in our universe, such 
as colliding neutron star binaries, supernova explosions and 
gravitational collapses into black holes. They manifest themselves 
as ripples in space and time, that periodically stretch and 
compress all present matter and delay and accelerate the time signals 
from millisecond pulsars, travelling to us over astronomical 
distances.

\subsection{Direct gravitational wave detection}
The only evidence, until now, of the existence of gravitational waves 
is the neutron star binary whose orbital radius is slowly 
decreasing as a consequence of the radiated gravitational energy. The 
reason that it is so hard to detect gravitational waves ({\sc gw}s) 
is that Newton's constant of gravitation is extremely small 
($G=6.7\cdot 10^{-11}\mbox{Nm}^2\mbox{kg}^{-2}$). The dimensionless 
amplitude ($\Delta l/l$) of typical gravitational waves reaching the 
earth is only of the order of $10^{-17}$. In other words, a rod of 
one meter in length will oscillate with an amplitude of a millionth 
of the radius of a hydrogen atom. 
Still, several direct {\sc gw}-detectors are being build at present, 
or have been proposed for the future, that hope to detect these very 
small vibrations. The best known of these are the {\em Laser 
Interferometer Ground Observatory}, {\sc ligo}, and the {\em Laser 
Interferometer Space Array}, {\sc lisa}. As the names suggest, the 
former is a terrestial observatory and the latter a space base one. 
Needless to say wha enormous engineering achievements these 
detectors require to detect the oscillations mentioned above.

Another approach to detect {\sc gw}s is to measure 
`relative time displacements' ($\Delta 
t/t$), instead of position displacements. Time signals from 
millisecond pulsars can be measured to such precision that time 
delays caused by the gravitational waves can in principle be measured 
with the same precision as position displacements as soon as enough 
stable pulsars can be observed for a sufficiently long time.
Dr. A. Lommen et. al. (Berkeley, U.S.A.) plan to use this fact to determine, from 
{\sc gw}s, whether the central object of our galaxy is a black 
hole binary.

\subsection{Watching gravitational waves}
The subject of the first part of this thesis is an entirely different 
way to {\em indirectly observe gravitational waves}. It appears that {\sc 
gw}s can be converted to electromagnetic waves, viz light, that are 
much easier to detect.

According to General Relativity, 
all forms of energy are equivalent, which is one way to formulate the 
Equivalence Principle. 
This is reflected by Einstein field equations, which 
are the equivalent of the non-relativistic Poisson equation for the 
gravitational potential. Just as 
the latter gives the gravitational potential as a function of the 
matter density, the former gives the space-time curvature due to the 
total energy-momentum density. Gravitational waves are the first 
order, harmonic solutions for these field equations.

As a result of the Equivalence Principle, not only oscillating matter 
sources can produce
{\sc gw}s, but also oscillating {\sc em} energy densities. 
In linearized theory, this means that also the reverse process might 
occur: electromagnetic waves generated by gravitational waves.
It will be shown in Sec. \ref{gerst} that a gravitational wave 
passing through a constant electromagnetic background field is 
partially converted into light.
In other words, one could indirectly `watch' gravitational waves in 
the form of light.

In Sec. \ref{gwemw} an exact calculation will be presented for a 
linearly polarized {\sc emw} passing through an arbitrary {\sc em} 
background field which converts it to a {\sc gw} which is then 
reconverted into an {\sc emw}. From this calculation it is 
apparent that such conversions are absolutely inefficient in any 
laboratory experiment. From an observational view point, however, they 
could provide a means to indirectly observe the {\sc gw}s from 
more energetic, astrophysical phenomena such as supernov{\ae} or 
merging neutronstar binaries.

\subsection{Cosmic background radiation}
As a second phenomenon where {\sc gw} to {\sc emw} conversions could 
play an important r\^ole, the observed fluctuations in the cosmic 
background radiation are investigated. According to the standard model big bang theories,
$20.000$ years after the Big Bang, the electron mist that until then obscured the universe, evaporated, and the 
universe became transparent to the $\approx 2700$ Kelvin thermal radiation.
 
This radiation is observed as a very isotropic flux of photons, coming to us from all directions. What has not been explained properly are the small fluctuations that do appear in the radiation. 
If, however, a primordial magnetic field existed in the early universe, gravitational waves travelling through this field could have been converted into electromagnetic waves causing small fluctuations in the otherwise homogeneous radiation.
An estimate of this effect is given in Sec. \ref{cbr}.

\subsection{Gamma ray bursts}
A final {\sc gw} to {\sc emw} conversion is discussed in the last 
part of this thesis. In a strongly magnetized plasma surrounding a 
source of {\sc gw}s, these waves might generate plasma waves, which 
in turn produce light in the form of cyclotron, synchrotron etc. 
radiation. In this case, the interaction is caused by the fact that 
in a very strong magnetic field, the plasma electrons are frozen to 
the field lines. The {\sc gw}s cause displacements of the electrons, 
which thus induce oscillations in the magnetic field. 

The relation to gamma ray bursts is that in these bursts, most of the 
energy is expected to be released in the form of gravitational waves, 
whereas the observed energy flux is electromagnetic.
A way to solve this problem would be to dissipate a small fraction of 
the gravitational wave energy into the plasma, which can then radiate 
this energy as the observed electromagnetic waves. This is a compelling 
alternative, for instance, to theories suggesting a neutrino fueled fireball.

\subsection{Overview}
The next section, Sec. \ref{arlt}, is a brief introduction to some of the concepts and equations that are needed in later sections.
In Sec. \ref{gws} gravitational waves are 
derived as weak field solutions of Einstein's field equations and the 
non-coordinate tedrad formalism, which will be used for most of the calculations, is explained in Sec. \ref{sectedrad}.

In Secs. \ref{gerst}--\ref{gwemw}, the efficiency of {\sc gw} to {\sc emw} 
conversion in a vacuum is considered, first in an estimate following 
Gertsenstein \cite{gertsenshtein} in Sec. \ref{gerst} and then, in an 
exact calculation, for an arbitrary {\sc em} background field and arbitrary 
{\sc gw} components in Sec. \ref{gwemw}.  
Finally, as an astrophysical application, 
an order of magnitude calculation is given for {\sc gw} to {\sc emw} conversion close to a 
quacking supernova remnant neutron star, or magnetar, in Sec. \ref{magnetar}.

In Sec. \ref{plasma} the vacuum is replaced by a thin plasma 
leading to some dispersion of the generated light (section \ref{wave}) and finally, in Sec. \ref{kuijp}, 
pressure gradients are added and 
the magnetohydrodynamic approximation is considered. From this 
the plasma waves are derived, needed to provide a {\sc gw} dissipation 
mechanism for gamma ray bursts.

The thesis ends with some conclusions in chapter \ref{conclusions} (and in appendix \ref{non-lin} second order {\sc gw}-effects are taken into account resulting in the excitation of longitudinal Alfv\'en-like waves).

\subsection{Note on units}
Throughout this thesis, Gaussian {\sc cgs} units are used in all equations. 
In most exact calculations the speed of light and Newton's constant are suppressed by choosing 
$c=G=1$ but in numerical estimates, these constants will sometimes reappear to emphasize their importance.

Furthermore, in the first two chapters greek indices will 
conventionally indicate four-vector components, $\mu = 0, 1, 2, 3$, 
with time as the zeroth component, and latin indices the spatial 
components $i,j = 1,2,3$. In the last chapter, however, the indices are used 
the other way around, following the particular convention used by the authors 
of the articles discussed in that chapter (\cite{brodin0600}-\cite{brodin0400} and 
\cite{papadopoulos1}-\cite{papadopoulos2}).

Finally, the signature of flat-space metric is chosen as:
\bq
\eta^{\mu\nu} =
\left(
\begin{array}{cccc}
-1 & 0 & 0 & 0 \\
0 & 1 & 0 & 0 \\
0 & 0 & 1 & 0 \\
0 & 0 & 0 & 1 
\end{array}
\right)
\eq
\newpage



\section{General relativity}\label{arlt}

\subsection{Gravitational waves}\label{gws}

\subsubsection{Introduction}
\begin{figure}[h!]
\begin{center}
\includegraphics[width=10cm]{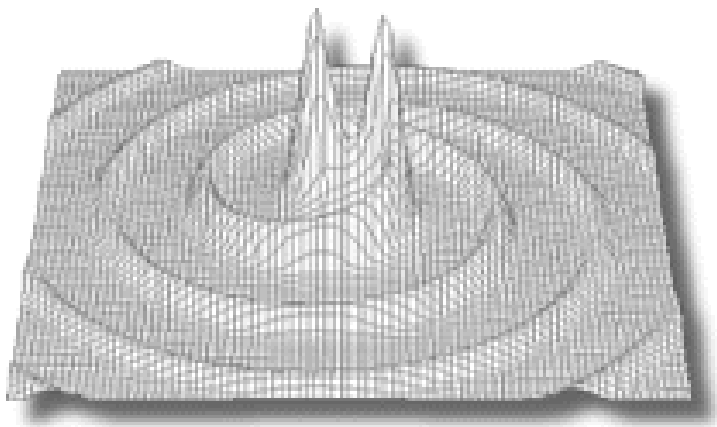}
\caption{\label{curvaturefig1}Gravitational wave from binary.}
\end{center}
\end{figure}

Gravitational waves, just like any other type of waves, are defined 
as propagating perturbations of some flat background.
Just as water waves are considered to be small ripples on an 
otherwise flat ocean, {\sc gw}s are identified as small ripples 
rolling across spacetime. And just as one neglects the deviation from 
flatness of the ocean caused by the curved surface of the earth, the tidal 
forces caused by the gravitational attraction of the sun and the 
moon, Coriolis forces due to the earths rotation etc., so one ignores 
the large-scale curved structure of space-time caused by the matter 
distribution or for instance the presence of a primordial magnetic 
background field. 
The {\sc gw}s originating from supernov{\ae}, explosions in the 
galactic center, or rotating binary stars can then indeed be described 
as small ripples on a flat background, as illustrated by the cartoon 
picture from {\sc lisa}, Figure \ref{curvaturefig1}, showing the 
gravitational waves of a binary.  

As a result of all this, the (linearized) wave equations are easy to 
derive, as will be done in this section. But one has to keep in mind 
that these equations are defined only locally, and that they have no 
meaning globally, where the large scale space-time structure of the 
universe comes into play.    

\subsubsection{Linearized theory in vacuum}
In the General Theory of Relativity, Poisson's equation of gravity is replaced by the equivalent, but covariant field equations, introduced by Einstein:
\bqa
\Box\Phi = 4\pi \rho &&\mdoos{\sc Poisson} \\
G_{\mu\nu} = R_{\mu\nu} - 
\half g_{\mu\nu} R = \frac{8\pi G}{c^4} T_{\mu\nu} &&\mdoos{\sc 
Einstein} \\\label{einstein} 
\mdoos{or} 
R_{\mu\nu} = \frac{8\pi G}{c^4} 
\haak{T_{\mu\nu} - \half g_{\mu\nu}T} &&
\eqa
The conceptual meaning of these equations is that a localized density distribution 
curves the space around it and as a result of this, the geodesics for 
light or particles are deflected in the direction of the center of 
mass, thus reproducing the effect of a gravitation force.

To derive the gravitational wave equations, the 
deviations from the Lorentz metric ($\eta_{\mu\nu}$) are assumed to be small. In 
other words, the weak-field solutions of Einstein's field equations 
(\ref{einstein}) are obtained by assuming:

\bqa\label{weakfield}
g_{\mu\nu} &=& \eta_{\mu\nu} + h_{\mu\nu} 
\mdoos{where} | h_{\mu\nu} | \ll 1 \\ \nonumber
h_{\mu}^{\ \lambda} &=& \eta^{\lambda\alpha} 
h_{\mu\alpha}\mdoos{and}h \equiv h_{\alpha}^{\ \alpha} = 
\eta^{\sigma\lambda}h_{\sigma\lambda}
\eqa
which is called the linearized theory of gravity.
In this metric the Ricci tensor to first order in $h_{\mu\nu}$ is 
(see for instance \cite{gravitation}):


\bqa\label{ricci}
R_{\mu\nu} &=& \Gamma^{\beta}_{\ \mu\nu,\beta} - 
\Gamma^{\beta}_{\ \mu\beta,\nu} +
\Gamma^{\beta}_{\ \mu\nu} \Gamma^{\alpha}_{\ \beta\alpha} -
\Gamma^{\alpha}_{\ \mu\beta}\Gamma^{\beta}_{\ \nu\alpha} \\ \nonumber	
&=&\Gamma^{\beta}_{\ \mu\nu,\beta} - \Gamma^{\beta}_{\ \mu\beta,\nu} 
\\ \nonumber	
&=&-\half h_{\mu\nu,}\,_{\alpha}^{\ \alpha} -
\half (h_{,\mu\nu} - h_{\mu}^{\ \beta}\,_{,\nu\beta} -	
h^{\beta}_{\ \nu,}\,_{\mu\beta}) \\ \nonumber	
&=&-\half h_{\mu\nu,}\,_{\alpha}^{\ \alpha} -
\half \haak{ 
(\half\eta_{\mu}^{\ \beta}h - h_{\mu}^{\ \beta})_{,\beta\nu} +
(\half\eta_{\nu}^{\ \beta}h - h_{\nu}^{\ \beta})_{,\beta\mu} } 
\\\nonumber
&=& -\half h_{\mu\nu,}\,_{\alpha}^{\ \alpha}
\eqa

\noindent where in the last line, the coordinate conditions are chosen such 
that $(h_{\mu}^{\ \beta} -\half\eta_{\mu}^{\ \beta}h)_{,\beta}=0$ 
(i.e. a divergence free and traceless solution). The field equations 
are now given (to first order) by:

\begin{eqnarray}\label{boxphi}
-\half h_{\mu\nu,}\,_{\sigma}^{\ \sigma} + 
\frac{1}{4}\eta_{\mu\nu} h_{,\sigma}^{\ \sigma} 
&=&\frac{8\pi G}{c^4} T_{\mu\nu} \mdoos{or} \\  \nonumber
\Box \phi_{\mu}^{\ \nu} &=& 
-\frac{16 \pi G}{c^4} T_{\mu}^{\ \nu} = 
-2 G^{(1)}\!_{\mu}\!^{\nu}\end{eqnarray}
where $\phi_{\mu}^{\ \nu}$ is defined by:

\bq\label{coor}
\phi_{\mu}^{\ \nu} = 
h_{\mu}^{\ \nu} - \half \eta_{\mu}^{\ \nu} h \mdoos{with}
\phi_{\mu}^{\ \nu}\,_{,\nu} =0 
\eq
The general solutions of (\ref{boxphi}) are, as mentioned in the 
previous section, just the solutions of the Poisson equation:
\bq\label{poiss}
\phi_{\mu}^{\ \nu} (r, t) = \frac{4G}{c^4} \int \frac{(T_{\mu}^{\ 
\nu})_{\mbox{\tiny ret}}d^3x^{\prime}}{|\vec{r}-\vec{r^{\prime}}|}
\eq

\subsubsection{Plane wave solutions}
In a vacuum and the absence of {\sc em} fields, where $T^{\mu\nu} = 0$, the field equations (\ref{boxphi}) reduces to 
$h_{\mu}^{\ \nu} - \half \eta_{\mu}^{\ \nu} h = 0$ with as simplest 
solutions plane waves:
\bqa
\phi_{\mu\nu} &=& \Re \brak{A_{\mu\nu}\mbox{e}^{ik_{\alpha} x^{\alpha}}} \mdoos{with} \\\nonumber
k_{\alpha}k^{\alpha} &=& 0 \mdoos{\sc null vector} \\\nonumber
A_{\mu\nu} k^{\nu} &=& 0 \mdoos{\sc transverse} 
\eqa
These solutions are transverse plane waves propagating with the speed of light ($\omega = 
|{\bf k}|$) in the direction ${\bf k}/\omega$.

The amplitude of this wave still seems to have six independent 
components (ten components because $A_{\mu\nu}$ is symmetric (since 
$T^{\mu\nu}$ is) of which four are fixed by the restriction 
$A_{\mu\nu} k^{\nu} = 0$).   
This is due to a gauge freedom under infinitesimal coordinate 
transformations $x^{\prime\mu} \rightarrow x^{\mu} + \xi^{\mu}$ 
satisfying $\Box \xi = 0$. 
In this case $\xi^{\mu} = -i C^{\mu} \exp{ik_{\alpha}x^{\alpha}}$ 
generates gauge transformations that can arbitrarily alter four of 
$A_{\mu\nu}$'s components.
therefore one has to choose not only a certain coordinate condition 
(as in (\ref{ricci})) but also a specific gauge.

\subsubsection{Transverse traceless gauge}
Defining the observer velocity by $u^{\nu}$, one can perform a 
gauge transformation such that $A_{\mu\nu}u^{\nu}=0$ which fixes 
three of the components, and another transformation to set 
$A^{\mu}_{\ \mu}=0$ wich fixes the last one. What remains are the two 
dynamical degrees of freedom of the gravitational field (the two 
polarizations).
Summarizing:
\bqa
A_{\mu\nu} u^{\nu} &=& A_{\mu\nu}k^{\nu} = A_{\ \mu}^{\mu} = 0 
\\\label{tt}
A_{\mu\nu} &=& 
\left(
\begin{array}{cccc}
0 & 0 & 0 & 0 \\
0 & A_+ & A_{\times} & 0 \\
0 & A_{\times} & -A_+ & 0  \\
0 & 0 & 0 & 0 
\end{array}\right)
\eqa
One result of this gauge is that $\phi_{\mu\nu} = h_{\mu\nu}$ which 
makes the analogy with Poisson's equation complete.
 
\subsubsection{Polarization}
Just like electromagnetic waves, all gravitational waves can be 
decomposed into two {\em linearly polarized} components, or in two 
{\em circularly polarized} components that give the geodesic 
deviation in a certain direction (\cite{gravitation}). 
Linearly polarized, electromagnetic waves with polarization vectors ${\bf 
e}_{x}$ or ${\bf e}_{y}$, travelling in the $z$-direction, cause a test 
particle to oscillates in the $x$- or $y$-direction respectively (with 
respect to an inertial frame).

For a gravitational wave, from (\ref{tt}), the {\em unit 
linear-polarization tensors} are, by analogy (see for instance 
\cite{jackson}):
\bqa
e_{+}^{\mu\nu} &\equiv & \frac{1}{\sqrt{2}} (e_{x}^{\mu} e_{x}^{\nu} 
- e_{y}^{\mu}e_{y}^{\nu} ) \\\nonumber 
e_{\times}^{\mu\nu} &\equiv & \frac{1}{\sqrt{2}} (e_{x}^{\mu} e_{y}^{\nu} 
+ e_{x}^{\nu}e_{y}^{\mu} ) 
\eqa
The effect of such a polarized wave on a ring of test particles around a 
central particle is as follows: in the plane transverse to the {\sc gw}-propagation,\footnote{in the 
reference frame of the central particle} the ring is deformed 
into an ellipse that pulsates in and out along the $x$- and $y$-axes 
(${\bf e}_{+}$) or at an angle of $45^{\circ}$ with respect to these axes 
(${\bf e}_{\times}$). In Sec. \ref{plasma}, plane $+$-polarized {\sc gw}s 
will be considered, entering a thin plasma. 

{\em Unit circular-polarization tensors} for gravitational waves are 
again constructed by analogy to electromagnetic polarization vectors, ${\bf 
e}_{R} = ({\bf e}_{x} + i{\bf e}_{y})/\sqrt{2}$ and  ${\bf 
e}_{L} = ({\bf e}_{x} - i{\bf e}_{y})/\sqrt{2}$:

\bqa
{\bf e}_{R} &=& \frac{1}{\sqrt{2}} ({\bf e}_{+} + i{\bf 
e}_{\times})\\\nonumber
{\bf e}_{L} &=& \frac{1}{\sqrt{2}} ({\bf e}_{+} - i{\bf e}_{\times}) 
\eqa
One can see that ${\bf e}_{R}$ and ${\bf e}_{L}$ rotate a deformation of a ring of test particles 
in the counterclockwise and the clockwise direction respectively. 

An interesting characteristic of these polarizations is that the 
gravitational wave is invariant under a $\pi$ rotation, 
which reflects the quantum behaviour of the associated massless {\em 
graviton} particles.

The classical radiation field of a spin $S$ particle is invariant 
under a $2\pi/S$ rotation (for instance $2\pi$ for spin-1 photons and 
$4\pi$ for spin-$\half$ neutrinos). Apparently, gravitons are 
spin-$2$ 
particles. This also follows from the fact that the two orthogonal 
linear polarizations of a spin-$S$ field are inclined to each other 
at an angle 
$\half \pi/S$. Indeed ${\bf e}_{+}$ and ${\bf e}_{\times}$ are 
inclined to each other at $\kwart \pi$ (and for instance for photons 
${\bf e}_{x}$ and ${\bf e}_{y}$ at $\half\pi$).

\subsubsection{Stress-energy carried by {\sc gw}}
A very important consequence of Einstein's equivalence principle is 
that 
{\em local gravitational energy-momentum}, or its density, is not 
uniquely defined. Non-vanishing stress-energy implies, after all, 
curved space, 
whereas the equivalence principle always allows one to choose a 
locally inertial (flat) coordinate system. Gravitational energy 
is therefore non-localizable. One has to average over several 
wavelengths to arrive at a well defined energy content in a certain 
region and talk about an {\em effective} smeared-out stress-energy of 
gravitational waves.

In an arbitrary gauge, this effective stress-energy for a {\sc gw} 
follows from contracting the weak field equations (\ref{einstein}) 
with $h^{{\mu}}_{\ \nu}\!_{,\rho}$ leading to (see \cite{weber} p. 94 
or \cite{gravitation} p. 955):

\bq\label{label}
T^{\mbox{\tiny \sc eff}}_{\mu\nu} = \frac{c^4}{16\pi G} \langle 
h_{\alpha\beta, \mu} h^{\alpha\beta}_{\ ,\nu} 
-\half h_{,\mu} h_{\nu} - 
h^{\alpha\beta}_{\ \beta} h_{\alpha\mu,\nu} 
 -  h^{\alpha\beta}_{\ \beta} h_{\alpha\nu,\mu} 
\rangle
\eq
which in the transverse-traceless gauge discussed before, reduces to:
\bq\label{t00}
T^{\mbox{\tiny \sc eff}}_{\mu\nu} = \frac{c^4}{16\pi G} \langle 
h^{TT}_{ij,\mu}h^{TT}_{ij,\nu}
\rangle
\eq
This is the expression that will be used in estimating the amount of 
gravitational energy that can be converted into electromagnetic 
radiation or vice versa when a {\sc gw} travels through a strong 
electromagnetic background field in for instance (\ref{too}) in Sec. 
\ref{gerst}.

For a plane, linearly polarized {\sc gw} of the form $h_{\mu\nu} = 
\Re\krul{ 
(A_{+}e_{+\mu\nu} + A_{\times}e_{\times\mu\nu})\mbox{e}^{ik(z-t)}}$, the only 
non-vanishing components are:
\bq
T^{\mbox{\tiny \sc eff}}_{tt} = T^{\mbox{\tiny \sc eff}}_{zz} =
- T^{\mbox{\tiny \sc eff}}_{tz} = - T^{\mbox{\tiny \sc eff}}_{tz} = \frac{k^{2}c^4}{16\pi 
G}(|A_{+}|^{2} + |A_{\times}|^{2})
\eq

\newpage

\subsection{Tedrad formalism}\label{sectedrad}

In describing physics on curved space-times, it is customary to work 
in the socalled tedrad formalism.
This formalism is usually the most efficient in computing curvature 
and many other quantities. Also it provides the closest connection to 
measurement. After all, a measurement made by a (for instance 
accelerated) observer is always made in its locally inertial 
reference frame. For this reason, most calculations in this thesis 
will, at some point, involve the introduction of a certain tedrad 
system to describe what physics is going on.
This section provides a short overview of the concepts that will be 
used in the following sections.

\subsubsection{Spinors and cosmological models} 
In the General Theory of Relativity, the only 
equations of physical interest are covariant tensor equations. 
Half integer spin particles, however, do not transform like tensors, 
but like spinors, so the general `recipe' of replacing all physical 
quantities by tensors, and using covariant derivatives everywhere 
instead of partial derivatives, does not work any more as soon as 
electrons or other spin $\half$ paricles are involved. It is 
therefore not possible to describe the effect of gravitation on the 
fields of electrons and most other charged particles (see 
\cite{weinberg}). Using non-coordinate methods, however, allows one 
to incorporate spinors into calculations on the same footing as 
tensors.

Another area where the tedrad formalism is of great importance is in 
the study of cosmological models. A lot of work on this is done by G. 
F. R. Ellis (\cite{ellis71}) and H. van Elst (in \cite{ellis98}). 
They set up these models in a $1+3$ covariant description where all 
tensor components are projected onto the average velocity vector and 
the directions orthogonal to it. The six propagation equations and 
six constraint equations resulting from this can only be completed by 
putting them into tedrad form, where one then has additional Ricci 
and Jacobi identities for the basis vectors.\footnote{The tedrad form 
also allows the set of equations to be put into a symmetric 
hyperbolic normal form which is easier to solve.}

Since the problems that will be discussed in this thesis involve a 
electron plasma in a cosmological background, it seems natural to 
adopt the tedrad formalism to describe these phenomena.

\subsubsection{Decomposition to locally inertial frame}
From the equivalence principle one is always free to choose a locally 
inertial coordinate system at every point $P$ (local metric 
$\eta_{\mu\nu}$), which is related to a general non-inertial system, $x^{{\mu}}$ by:
\bqa
g_{\mu\nu} (x) &=& 
\haak{ \frac{\partial \xi_{P}^{\alpha} (x)}{\partial x^{\mu}}}_{x=P} 
\haak{ \frac{\partial \xi_{P}^{\beta} (x)}{\partial x^{\nu}}}_{x=P} 
\eta_{\alpha\beta} \\\nonumber
&=& \lambda^{\alpha}_{\ \mu}\lambda^{\beta}_{\ \nu}\eta_{\alpha\beta}
\eqa
One can easily see any other choice of general coordinates 
$x^{\prime\mu}$ would have led to the same expression, with:
\bqa
\lambda^{\prime\alpha}_{\ \mu} &=&
\frac{\partial \xi_{P}^{\alpha}}{\partial 
x^{\prime\mu}} =
\frac{\partial x^{\nu}}{\partial x^{\prime\mu}}
\frac{\partial \xi_{P}^{\alpha}}{\partial x^{\nu}} =  
\frac{\partial x^{\nu}}{\partial x^{\prime\mu}}
\lambda^{\alpha}_{\ \nu} 
\eqa
Apparently, $\lambda^{\alpha}_{\ \mu}$ does not transform like a 
single 
tensor, but as four covariant vector fields, hence the name {\em 
tedrad} 
or {\em vierbein}.
Using this tedrad, any contravariant vector field or tensor can be 
decomposed along the locally inertial coordinate axes:
\bqa
F^{(\mu)} &\equiv& \lambda^{(\mu)}_{\ \nu}F^{\nu} \\\nonumber
G^{(\mu\nu)} &\equiv & \lambda^{(\mu)}_{\ \alpha} \lambda^{(\nu)}_{\ 
\beta}  G^{\alpha\beta}
\eqa
where the brackets denote which tedrad vector is contracted, and 
illustrate that by contracting the tensor fields with the tedrad 
vectors, they are decomposed into a set of scalars.\footnote{Later on 
these brackets will be omitted for brevity, assuming that it is clear 
to the reader what the different indices mean.} 
This is precisely the reason why the treatment of spinors (of for instance Dirac 
electrons) is no 
longer different from that of tensors. Both can be put into an 
action as scalars.

In a more informal way of speaking, what is meant by the tedrad decomposition, is that 
at each point in space, physical quantities are projected onto artificial, orthogonal 
axis that `stick out' of the curved space as it where. Hence the name {\em non-coordinate} frames.

\subsubsection{General covariance and Lorentz invariance}

To construct an action from the decomposed tensors, both general 
covariance and Lorentz 
invariance of the action must be satisfied.
The scalar components are, after all, defined with respect to an 
arbitrarily 
chosen {\sc onf}, so the action should be invariant under a 
redefinition of this frame, viz under Lorentz transformations.
\bqa\label{lorentztrans}
F^{(\mu)} (x) &=& \Lambda^{\mu}_{\ \nu} (x) F^{(\nu)} (x) \\\nonumber
G^{(\mu\nu)} &= & \Lambda^{\mu}_{\ \alpha} (x) \Lambda^{\nu}_{\ 
\beta} (x) G^{(\alpha\beta)} (x)
\eqa
The same conditions apply to the tedrad vectors themselves. 

Now, to construct an appropriate action, one does not only need the 
fields, decomposed as scalars, but also derivatives of the fields. 
Since al the fields are decomposed as scalars, the derivatives also 
have to be scalars. At the same time though they have to be invariant 
under Lorentz transformations. The derivative that satisfies 
these conditions is (see \cite{ellis98}, \cite{stewart} or 
\cite{weinberg}) :
\bqa\label{ricciafg}
\nabla_{\mu} &=& 
\lambda_{(\mu)}^{\ \nu}\frac{\partial}{\partial x^{\nu}} - 
\lambda^{(\alpha)}_{\ \beta} \lambda_{(\mu)}^{\ \gamma} 
\frac{\partial}{\partial x^{\gamma}} \lambda_{(\delta)}^{\ \beta}
\eqa
\paragraph{Summarizing:}
just as one obtains General Relativity by replacing the special 
relativistic action and field equation by tensor equations, and the 
partial derivatives by covariant deriveves, one obtains a coordinate 
free description of General Relativity by decomposing all tensors or 
spinors into 
scalars (except for the tedrad vectors itself), and replacing all the 
partial derivatives with the covariant and Lorentz invariant 
derivative 
(\ref{ricciafg}).

\subsubsection{The connection}
In a particular set of coordinates, space-time is described by a 
metric given by $g^{\mu\nu}(x)$ and a connection, that supplies the 
differential properties of space-time: the Christoffel symbols. In 
the non-coordinate formalism, space-time is is locally flat (metric 
$\eta^{\mu\nu}$) with respect to a particular set of tedrad vectors, 
and its differential properties are given by a connection in the form 
of the socalled {\em Ricci rotation coefficients}.

The specific form of these coefficients follows readily from 
(\ref{ricciafg}).      
\bqa
\nabla_{\bf \lambda_{\beta}} {\bf \lambda_{\alpha}} &=& 
\ric{\gamma}{\alpha\beta} {\bf \lambda_{\gamma}} 
\mdoos{$\Leftrightarrow $}\\\nonumber 
\ric{\alpha}{\beta\mu} &=& 
\lambda^{(\alpha)}_{\ \beta} \lambda_{(\mu)}^{\ \gamma} 
\frac{\partial}{\partial x^{\gamma}} \lambda_{(\delta)}^{\ \beta}
\eqa
In tedrad components, covariant derivatives can thus be calculated 
completely analogous to the tensorial form:
\bq
\nabla_{\alpha} T_{\beta\gamma} = 
\lambda_{(\alpha)}^{\ \nu}\frac{\partial T_{\beta\gamma}}{\partial 
x^{\nu}} - 
\ric{\delta}{\beta\alpha} T_{\delta\gamma} - 
\ric{\delta}{\gamma\alpha} T_{\beta\delta}
\eq
This derivative will be used in all the plasma calculations is Sec. 
\ref{plasma}.

\newpage

\section{Estimate of interaction efficiency}\label{gerst}

According to General Relativity, gravitational waves and 
electromagnetic waves propagate with the same speed and, in vacuum, 
obey the same dispersion relation. If there is a 
linear dependence between the waves, they can therefore resonate and 
transfer energy (\cite{gertsenshtein}). In this section propagation 
of light ($F^{\mu\nu}$) in the presence of a strong magnetic 
background field ($F^{(0)}\!^{\mu\nu}$) that is constant in 
space-time, is considered. 

The stress-energy tensor for the combined field 
($F^{\prime}\!^{\mu\nu} = F^{(0)}\!^{\mu\nu} + F^{\mu\nu}$) consists 
of:
\begin{enumerate}
\item the square of the constant field term that does not generate 
gravitational waves, 
\item the square of the field of the electromagnetic wave which does not create {\sc 
gw}'s either, and
\item the interference term which is proportional to the background 
field and produces an oscillating source term for gravitational waves.
\end{enumerate}
The field equations for the resonant term, which is the only relevant 
part of the stress-energy tensor, are given by:

\bqa\label{resonance}
\Box\phi^{\mu}_{\ \nu} &=& -\frac{8G}{c^4} 
\brak{
F^{(0)}\!^{\mu\alpha}F_{\nu\alpha} - 
\frac{1}{4} \eta^{\mu}_{\ \nu}	
F^{(0)}\!^{\alpha\beta}F_{\alpha\beta}     
} \\\nonumber
&=& -\frac{8G}{c^4} 
F^{(0)}\!^{\mu\alpha}F_{\nu\alpha}
\eqa
where the second line results from choosing certain convenient coordinates, for instance such that:
\bqa\nonumber
F^{(0)}\!^{\alpha\beta}F_{\alpha\beta} &\propto & ({\bf E} \cdot {\bf 
E}^{(0)} - {\bf B}\cdot{\bf B}^{(0)}) 
\mdoos{with} \\\label{pola}
{\bf E}^{(0)} &=& 0 \mdoos{and}  {\bf B} \bot {\bf B}^{(0)}
\eqa

For an incoming lightwave along the $x$-axis (wavevector 
$|\vec{k}|=k_x=\omega/c$), with no absorption or scattering ($b(x)=b$), 
try plane wave solutions with amplitudes normalized to unit energy 
density (\ref{t00}) and $k^{\alpha}_{\mbox{\tiny \sc 
gw}}=k^{\alpha}_{\mbox{\tiny \sc emw}}=k^{\alpha}$, e.g. 
gravitational waves produced along the $x$-axis. 

\begin{figure}[h!]
\begin{center}
\includegraphics[width=12cm]{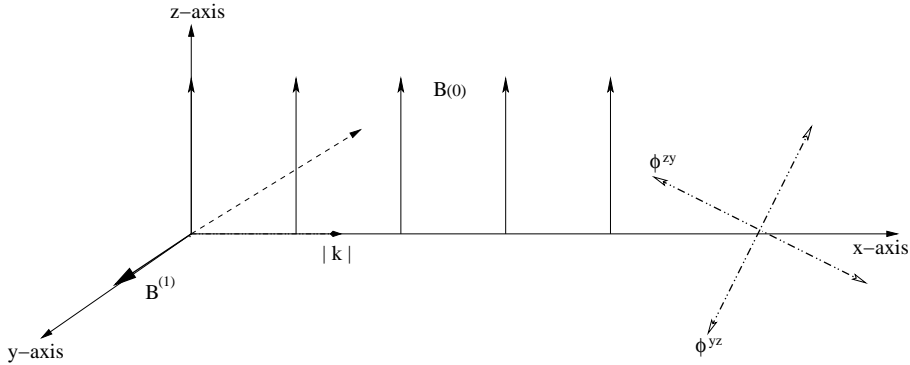}
\caption{\label{gertfig1} Orthogonal magnetic fields.}
\end{center}
\end{figure}

\begin{figure}[h!]
\begin{center}
\includegraphics[width=12cm]{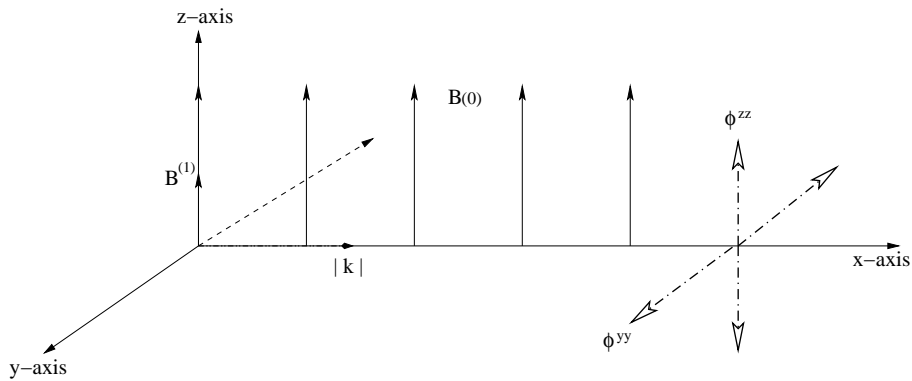}
\caption{\label{gertfig2} Parallel magnetic fields.}
\end{center}
\end{figure}

\bqa\label{too}
t_{00} &\sim & \frac{c^4}{16\pi G}<(h_{\mu\nu,0})^2> = 
\frac{c^4}{16\pi G}\langle \phi_{\mu\nu,0}\rangle^2 \\\nonumber
T_{00} &=& \frac{1}{4\pi}({\bf E}^2 + {\bf B}^2)
\eqa 
except for the overall `conversion factors' $a(x)$ and $b$:
\bqa\label{solution}
F_{\mu\nu} &=&  
\Re \brak{ b \sqrt{4\pi}f_{\mu\nu} \mbox{e}^{ik_{\alpha}x^{\alpha}}} 
\mdoos{,} f^{0\nu}f_{0\nu} = 1 \mdoos{and} f^{ij}f_{ij}=1\\\nonumber
\phi^{\mu\nu} &=& 
\Re \brak{a(x) \sqrt{\frac{16\pi G}{c^4 k^2}} \zeta^{\mu\nu} 
\mbox{e}^{ik_{\alpha}x^{\alpha}}} 
\mdoosje{,} \zeta^{\mu\nu} \zeta_{\mu\nu}  = 1 
\mdoosje{,} \zeta^{\mu}_{\ \mu} = 0 
\eqa
For slowly varying amplitudes, neglecting terms quadratic in $d/dx$ 
and of course $k_{\mu}k^{\mu}=0$, the field equations reduce to:

\bqa\label{ax}
\Box \phi^{\mu}_{\ \nu} &=&\sqrt{\frac{16\pi 
G}{c^4}}\frac{\zeta^{\mu}_{\ \nu}}{k_{x}}
\haak{k_{\alpha}k^{\alpha} a(x) +\frac{d^2 a(x)}{dx^2} + 
2ik_{x}\frac{d a(x)}{dx} }\mbox{e}^{ik_{\beta}x^{\beta}} 
\\\nonumber
&=& 
-\frac{16 \sqrt{\pi} bG}{c^4}F^{(0)\mu\alpha}f_{\nu\alpha}\mbox{e}^{ik_{\beta}x^{\beta}}
\eqa
which is a differential equation that can be solved for $a(x)$:
\bqa
\frac{d a(x)}{dx} &=&ib \sqrt{\frac{4G}{c^4}} 
F^{(0)\mu\alpha}f_{\nu\alpha} \zeta_{\mu}^{\ \nu} \\ \label{test}
a(x) &=& i b\sqrt{\frac{4G}{c^4}} f_{\nu\alpha} \zeta_{\mu}^{\ 
\nu} \int_{0}^{x} F^{(0)\mu\alpha} (x^{\prime}) dx^{\prime} + a(0)
\eqa
One can already see from this that the energy of the generated {\sc gw} grows with 
distance ($a^{\ast} a \uparrow x$). In the figures below, the polarization of the generated {\sc gw} is shown for 
two different polarizations for the incident {\sc emw}. In Figure \ref{gertfig1}, the magnetic wave component is orthogonal to the magnetic background field, as in (\ref{pola}): ${\bf B}_{\mbox{\tiny \sc em}}\bot {\bf B}^{(0)} \bot {\bf k}_{\mbox{\tiny \sc em}} $, and the resulting {\sc gw} is $\times$-polarized.
\newpage
\noindent
One could of course also choose the incoming {\sc emw} to have a magnetic component parallel to the background field. A similar calculation shows that for ${\bf B}_{\mbox{\tiny \sc em}}  \| {\bf B}^{(0)} \bot {\bf k}_{\mbox{\tiny \sc em}}$, the resulting {\sc gw} is $+$-polarized. This is depicted in Figure \ref{gertfig2}.

\noindent
Linear combinations of these electromagnetic configurations lead to circularly polarized {\sc gw}s (or any other polarization).
 
\paragraph{Efficiency.}

\noindent
To calculate the efficiency, assume:
\begin{itemize}
\item[$\blacklozenge $] a static, homogeneous magnetic background field, 
$F^{(0)}$, 
\item[$\blacklozenge $] a typical length- or timescale for the interaction region, 
$L=Tc$, 
\item[$\blacklozenge $] unit convolutions of dimensionless amplitudes, 
\item[$\blacklozenge $] and no incoming gravitational waves, $a(0)=0$. 
\end{itemize}
\noindent
The energy transfer efficiency from {\sc emw}s to {\sc gw}s is then given by:
\bq\label{gersteffi}
\alpha = \Bigg\|\frac{a(x)}{b}\Bigg\|^2 = \frac{4 G}{c^4} F^{(0) 2} L^2
\eq
This efficiency is the same for {\sc emw} $\Rightarrow$ {\sc gw} and 
{\sc gw} $\Rightarrow $ {\sc emw} conversions since the relations are symmetric
under time reversal.
For a neutronstar binary or magnetar with a large surface magnetic 
field $F^{(0)} \approx 10^{16}$ Gauss and an 
interaction region (where one can speak of plane {\sc gw}s) from 
$R_1 = 180$ km to $R_2 = 500$ km, and a 
dipolar decay of the magnetic field, something of the order of 
$10^{-8}$ of the energy could be converted, which still might be 
substantial considering the huge amounts of energy released in 
supernov{\ae} and binary mergers. 
  
\paragraph{Fluctuations in the {\sc cbr}.}

A very interesting example of {\sc gw} to {\sc emw} conversions, could have occured 
in the early universe. The observed fluctuations of in the cosmic background radiation 
might be explained by gravitational waves travelling throught the magnetic field that, according to some
cosmological models existed throughout space at that time. 
If a fraction of the {\sc gw} energy could be converted into {\sc emw}s, this could explain the $\sim 10^{-5}$
relative anomalies in the {\sc gbr}. This will be discussed in some more detail in Sec. \ref{conclusions}.

\newpage
\section{Exact calculation in vacuum.}\label{gwemw}

Motivated by the result of the preceding crude estimate, this section 
proceeds to the exact, analytical derivation of a conversion of 
photons into gravitons and back into photons again. 

\subsection{Photons to gravitons} 

\begin{figure}[h!]
\begin{center}
\includegraphics[width=12cm]{%
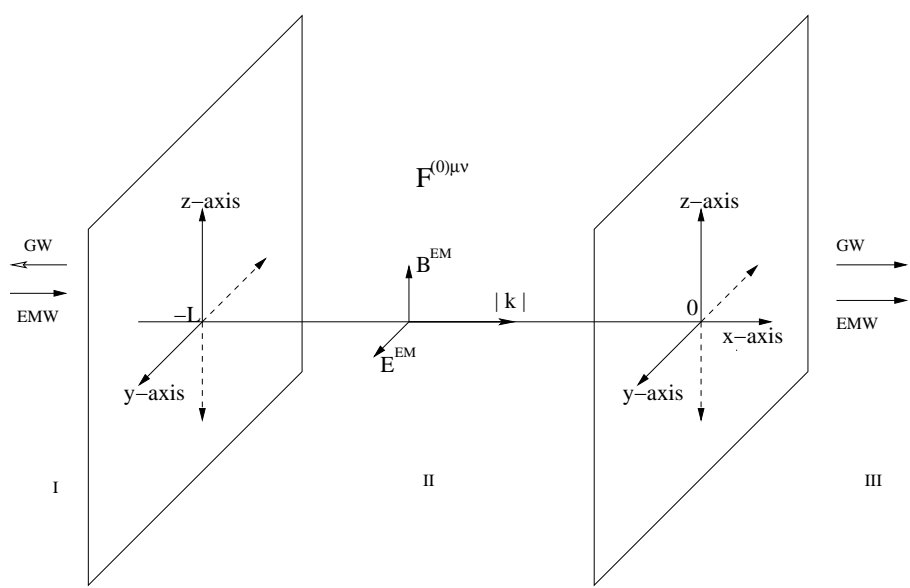}
\caption{\label{fig:1}Photon to graviton conversion.}
\end{center}
\end{figure}

Consider the arrangement shown in Figure \ref{fig:1} (\cite{lupanov} 
and \cite{boccaletti}), where an electromagnetic wave is partially 
converted into a gravitational wave. 
The coordinates are again chosen such that the incident electromagnetic wave 
propagates along the positive $x$-axes. Furthermore, the waves are 
taken to be linearly polarized plane waves: $B_{z} = E_y = A\exp 
ik(x-ct)$. For a completely general electromagnetic background field 
the combined Faraday tensor is:

\bq\label{tem}
F_{\mu\nu} =
\left(
\begin{array}{cccc}
0 				   & E_x 		& E_y + A\mbox{e}^{ik(x-ct)} & E_z \\
-E_x 				   & 0		& B_{z} + A\mbox{e}^{ik(x-ct)} & -B_{y} \\
-E_y - A\mbox{e}^{ik(x-ct)}& - B_{z} - A\mbox{e}^{ik(x-ct)} & 0 		& B_{x} \\
-E_z				   & B_{y}					& -B_{x}     & 0 \\
\end{array}
\right)
\eq
As in Sec. \ref{gerst}, the stress-energy tensor consists of two 
squared terms, that do not produce gravitational waves, and the 
resonant term of interest, which in this case is:


\bq
T_{\mu}^{\ \nu} = \frac{A}{4\pi} 
\left(
\begin{array}{cccc}
-(E_y + B_{z})	& -(E_y + B_{z}) 	& E_x  		& B_{x} \\
(E_y + B_{z}) 	& (B_{z} + E_y)	& E_x  		& B_{x} \\
-E_x	    	& -E_x	  	& (B_{z} - E_y)	& -(B_{y} + E_z) 
\\-B_{x}	  	& -B_{x}	  	& -(B_{y} + E_z)	& -(B_{z} - E_y) \end{array}\right)
\mbox{e}^{ik(x-ct)}
\eq
\noindent
satisfying the usual conservation condition $T_{\mu}^{\ \nu}\!_{;\nu} 
= 0$. So for the ten independent components of the stress-energy 
tensor, the Einstein equations (\ref{boxphi}) read: 


\bqa\label{einsteinx}
\Box \phi_{I\mu}^{\quad\nu} &=& 0  \\\nonumber
\Box \phi_{II\mu}^{\quad\nu} &=& \zeta_{\mu}^{\ \nu} 
\mbox{e}^{ik(x-ct)} \\\nonumber
\Box \phi_{III\mu}^{\quad\nu}&=& 0
\eqa
\noindent
where the subscripts $I$, $II$ and $III$ refer to the regions as 
indicated in Figure \ref{fig:1}. The amplitude $\zeta_{\mu}^{\ \nu}$ 
is defined by: 


\bq\label{zeta}
\zeta^{\mu\nu} = 
-\frac{4AG}{c^4} 
\left(
\begin{array}{cccc}
-(E_y + B_{z})& -(E_y + B_{z}) & E_x  		& B_{x} \\
-(E_y + B_{z}) & (E_y + B_{z}) & -E_x  		& -B_{x} \\
E_x	    	& -E_x	  & (B_{z} - E_y)		& -(B_{y} + E_z) \\
B_{x}	  	& -B_{x}	  & -(B_{y} + E_z)	& -(B_{z} - E_y) \\
\end{array}
\right)
\eq
The general plane wave solutions satisfying (\ref{einsteinx}) are:


\bqa
\phi_{I\mu}^{\quad \nu} &=& \zeta_{\mu}^{\ \nu} A_{(\mu\nu )} 
\mbox{e}^{-ik(x+ct)}  \\\phi_{II\mu}^{\quad \nu} &=& \zeta_{\mu}^{\ 
\nu}\haak{\frac{x+L}{2ik}\mbox{e}^{ik(x-ct)} + B_{(\mu\nu )} 
\mbox{e}^{ik(x-ct)} + C_{(\mu\nu )} \mbox{e}^{-ik(x+ct)}} 
\\\phi_{III\mu}^{\quad \nu} &=& \zeta_{\mu}^{\ \nu} D_{(\mu\nu )} 
\mbox{e}^{ik(x-ct)} 
\eqa
where the arbitrary constants $A_{(\mu\nu )}, B_{(\mu\nu )}, 
C_{(\mu\nu )}, D_{(\mu\nu )}$ still have to be determined from the 
coordinate conditions ($\phi_{\mu}^{\ \nu}\!_{,\nu} = 0$ as in 
(\ref{coor})) and the continuity conditions at $x=-L$ and $x=0$:


\bqa
\phi_{I\mu}^{\quad\nu} |_{x=-L} &=& \phi_{II \mu}^{\quad\nu} |_{x=-L} 
\\\phi_{II\mu}^{\quad\nu} |_{x=0} &=& \phi_{III \mu}^{\quad\nu} 
|_{x=0} \mdoos{and} \\\phi_{I\mu}^{\quad\nu}\!_{,1} |_{x=-L} &=& 
\phi_{II \mu}^{\quad\nu}\!_{,1} |_{x=-L} 
\\\phi_{II\mu}^{\quad\nu}\!_{,1} |_{x=0} &=& \phi_{III 
\mu}^{\quad\nu}\!_{,1} |_{x=0} \eqa
Some tedious algebra results in the following constants:


\bqa
A_{(\mu\nu )} &=& 
\frac{1}{4k^2}
\left(\begin{array}{rrrr}
1-\mbox{e}^{2ikL} & \mbox{e}^{-2ikL}-1 &  \mbox{e}^{-2ikL} -1 &  
\mbox{e}^{-2ikL} -1 \\
1-\mbox{e}^{2ikL} & \mbox{e}^{-2ikL}-1 & -\mbox{e}^{2ikL} +1 &   
-\mbox{e}^{2ikL} +1\\
1-\mbox{e}^{2ikL} & \mbox{e}^{-2ikL}-1 & \mbox{e}^{-2ikL} -1 & 
\mbox{e}^{-2ikL} -1 \\
1-\mbox{e}^{2ikL} & \mbox{e}^{-2ikL}-1 & -\mbox{e}^{2ikL} +1  & 
\mbox{e}^{-2ikL} -1
\end{array}\right) 
\\ \nonumber
B_{(\mu\nu )} &=& \frac{1}{4k^2}\left(
\begin{array}{rrrr}
-1 & 1 &  1 & 1 \\
-1 & 1 & -1 &-1 \\
-1 & 1 &  1 & 1 \\
-1 & 1 & -1 & 1 
\end{array}
\right) \\ \nonumber
C_{(\mu\nu )} &=& \frac{1}{4k^2}\left(
\begin{array}{rrrr}
1 & -1 & -1 &  1 \\
1 & -1 &  1 &  1 \\
1 & -1 & -1 & -1 \\
1 & -1 &  1 & -1 
\end{array}\right) \mdoosje{,}
D_{(\mu\nu )} = \frac{L}{2ik}\left(
\begin{array}{rrrr}
1 & 1 & 1 & 1 \\
1 & 1 & 1 & 1 \\
1 & 1 & 1 & 1 \\
1 & 1 & 1 & 1
\end{array}\right)
\eqa
In these equations the indices in brackets denote the constant factors of 
the corresponding index positions in $\zeta_{\mu}^{\nu}$.

If the length of the device is set to a half-integer multiple of the 
wavelength, the amplitude of the retreating gravitational waves 
vanishes (for $L=n\lambda /2 \rightarrow \exp(\pm 2ikL)=1$). In this 
case gravitational waves are only generated in region III, which are 
linearly dependent on the length of the static background field. The 
complete expression for these waves in terms of the controlled 
parameters (background field, incident {\sc em}-waves and length of 
the device) becomes:


\bqa\label{supersolution1}
\phi_{\mu}^{\ \nu} &=& 
\frac{L}{2ik}\zeta_{\mu}^{\ \nu}\mbox{e}^{ik(x-ct)} \\ \nonumber
&=& \xi_{\mu}^{\ \nu} \mbox{e}^{ik(x-ct)} = h_{\mu}^{\ \nu}
\eqa
The last line follows from: $h_{\mu}^{\ \nu} = \phi_{\mu}^{\ \nu} - 
\half \eta_{\mu}^{\ \nu} \phi_{\alpha}^{\ \alpha}$ with, from 
(\ref{zeta}), $\phi_{\alpha}^{\ \alpha} =0$. 
As a result of this, (\ref{supersolution1}) fully describes the 
metric of the {\sc gw} entering the next part of the detection device 
(Sec. \ref{gravphot}).

\subsubsection{Efficiency}\label{label1}
From the equation for the metric (\ref{supersolution1}), one can obtain an expression for 
the energy flux (\ref{label}) carried by the generated gravitational wave:


\bqa\label{effi2}
W = t^{01} &=& \frac{c^{2}}{16\pi G} \haak{\dot{h}_{23}^{2}+ 
(\dot{h}^{2}_{22}-\dot{h}^{2}_{33})^{2}} \\ \nonumber
&=&\haak{\frac{G}{4\pi c^4}} (LA)^2 \haak{(B_{y}+E_z)^2 + (E_y + B_{z})^2}
\eqa
Apparently, there is no coupling when a linearly polarized {\sc emw} 
with components $B_z=E_y$, propagates through a background field with $E_y = - B_z$ 
and $E_z= - B_y$. Because of symmetry arguments, the same result will hold for an incident 
linearly polarized {\sc emw} with $B_y=E_z$.
For a pure magnetic field (i.e. $F^{\mu\nu} \rightarrow B_{z}$) this 
reduces to the estimated relation derived in the previous section, 
except for an overall factor of $1/16\pi$, whereas for a pure magnetic or 
electric field aligned with the $x$-axis, the energy flux vanishes! 
Note the quadratic dependence of the energy flux on both the length 
of the static field and the amplitude of the incident {\sc em}-wave.

\newpage

\subsection{Gravitons to photons}\label{gravphot}

\begin{figure}[h!]
\begin{center}\includegraphics[width=12cm]{%
    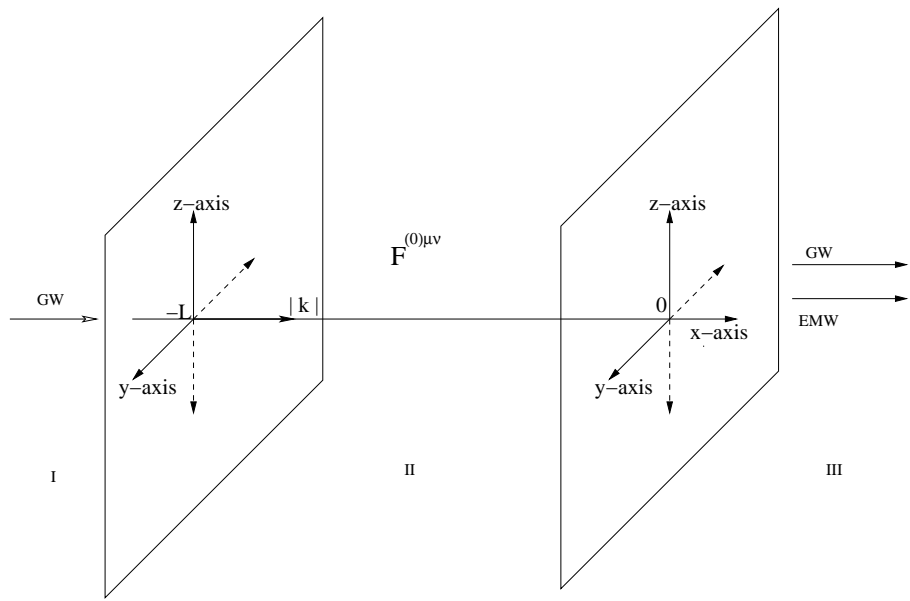}

\caption{\label{fig:2}Graviton to photon conversion.}
\end{center}
\end{figure}
Let us proceed now to the second part of the detection device, where 
the incident gravitational waves are reconverted to electromagnetic 
waves (Figure \ref{fig:2}).\footnote{After travelling through the first background field, 
the remaining light is blocked by a screen that does not influence 
the propagation of the {\sc gw}s.} The same 
symmetry relative to the $x$-axis as before is used to solve the 
Maxwell equations,


\bq
F_{\alpha\beta ,\gamma} + F_{\gamma\alpha , \beta} + F_{\beta\gamma 
,\alpha} = 0\mdoos{and}(\sqrt{-g}F^{\mu\nu})_{,\nu} = 
F^{\mu\nu}\!_{,\nu} = 0
\eq
in the oscillating metric perturbation $h_{\mu\nu}$. Because of the symmetry, only 
the derivatives with respect to $x$ and $ct$ contribute (to first order, these are just partial derivatives, since the connection is itself proportional to $h_{\mu\nu}$):


\bqa\label{maxwellfout}
F_{23,1} = F_{23,0}=0 &\mdoosje{,}& F^{01}\!_{,1} = F^{10}\!_{,0} =0 
\\\nonumber
F_{13,0} + F_{30,1}=0 &\mdoosje{,}& F^{31}\!_{,1} + F^{30}\!_{,0} =0 
\\\nonumber
F_{21,0} + F_{02,1}=0 &\mdoosje{,}& F^{21}\!_{,1} + F^{20}\!_{,0} =0 
\eqa
which results in the first field components $F_{23} = F_{23}^{(0)}$ 
and $F^{01} = E^{(0)}_{x}$. In the specified metric, the 
contravariant components for $F^{\mu\nu}$ are derived by:


\bqa
F^{\mu\nu} &=& g^{\mu\alpha} g^{\nu\beta} F_{\alpha\beta} \\\nonumber
&=& (\eta^{\mu\alpha} + h^{\mu\alpha})(\eta^{\nu\beta} + 
h^{\nu\beta})F_{\alpha\beta} \\\nonumber
&=& (\eta^{\mu\alpha}\eta^{\nu\beta} + \eta^{\mu\alpha}h^{\nu\beta} + 
h^{\mu\alpha}\eta^{\nu\beta})F_{\alpha\beta} + {\cal O}(h^2)
\eqa
so for $F^{23}$:


\bq
F^{23} = B_{x}^{(0)} + (E_y^{(0)} - B_{z}^{(0)})h_{30} + (B_{y}^{(0)} + 
E_z^{(0)})h_{12}
\eq
Similarly, the contravariant components of the other covariant 
differential equations are given by (using the symmetries in the 
metric tensor): 


\bqa\label{maxwell}
F^{21,0} + F^{02,1} &=& (E_y^{(0)} - B_{z}^{(0)})h_{22,1} + (B_{y}^{(0)} 
+ E_z^{(0)})h_{23,1} \\\nonumber
F^{13,0} + F^{30,1} &=& (E_y^{(0)} - B_{z}^{(0)})h_{23,1} + (B_{y}^{(0)} 
+ E_z^{(0)})h_{33,1}
\eqa
which together with (\ref{maxwellfout}) leads to the wave equations:


\bqa\label{boxfaraday}
\Box F^{21} &=& \\\nonumber 
\Box F^{20} &=& (E_y^{(0)} - B_{z}^{(0)})h_{22,11} + (B_{y}^{(0)} + 
E_z^{(0)})h_{23,11} \\\nonumber
&=& \alpha \mbox{e}^{ik(x-ct)} \\\nonumber
\Box F^{31} &=& \\\nonumber 
\Box F^{30} &=& (E_y^{(0)} - B_{z}^{(0)})h_{23,11} + (B_{y}^{(0)} + 
E_z^{(0)})h_{33,11} \\\nonumber
&=& \beta \mbox{e}^{ik(x-ct)}
\eqa
with,

\bqa\label{boxfaraday2}
\alpha &=& -k^2 \krul{(E_y^{(0)} - B_{z}^{(0)})\xi_{22} + (B_{y}^{(0)} + 
E_z^{(0)})\xi_{23}}\\\nonumber
\beta &=&  -k^2 \krul{(E_y^{(0)} - B_{z}^{(0)})\xi_{23} + (B_{y}^{(0)} + 
E_z^{(0)})\xi_{33}}
\eqa
In the same fashion, one can express (\ref{maxwell}) in 
$\alpha\mbox{e}^{ik(x-ct)}/ik$ and $\beta\mbox{e}^{ik(x-ct)}/ik$.
The general solutions of (\ref{maxwellfout}), (\ref{maxwell}) and 
(\ref{boxfaraday}) are just in- and outgoing plane wave solutions 
plus a linear term expressing how the {\sc emw}s grow in amplitude 
while the {\sc gw} loses energy. The electromagnetic 
field components are given by the solutions: 


\bqa
B_{z} &=& B_{z}^{(0)} + \frac{\alpha x}{2ik}\mbox{e}^{ik(x-ct)} + A 
\mbox{e}^{ik(x-ct)} + B \mbox{e}^{-ik(x+ct)} \\ \nonumber
-E_y &=& -E_{y}^{(0)} + \haak{\frac{\alpha x}{2ik} - 
\frac{\alpha}{2k^2}}\mbox{e}^{ik(x-ct)} + A \mbox{e}^{ik(x-ct)} - B 
\mbox{e}^{-ik(x+ct)} \\ \nonumber
B_{y} &=& H_{y}^{(0)} + \frac{\beta x}{2ik}\mbox{e}^{ik(x-ct)} + C 
\mbox{e}^{ik(x-ct)} + D \mbox{e}^{-ik(x+ct)} \\ \nonumber
-E_z &=& -E_{z}^{(0)} + \haak{\frac{\beta x}{2ik} - 
\frac{\beta}{2k^2}}\mbox{e}^{ik(x-ct)} + C \mbox{e}^{ik(x-ct)} - D 
\mbox{e}^{-ik(x+ct)} 
\eqa
where the arbitrary constants $A, B, C$ and $D$ still have to be 
determined from continuity conditions on the oscillating field 
($i=x,y,z$):


\bqa
E_{Ii}|_{x=-L} = E_{IIi} |_{x=-L} &\mdoos{and}& B_{Ii}|_{x=-L} = 
B_{IIi}|_{x=-L} \\ \nonumber
E_{IIi}|_{x=0} = E_{IIIi} |_{x=0} &\mdoos{and}& B_{IIi}|_{x=0} = 
B_{IIIi}|_{x=0} 
\eqa

\subsubsection{Tedrad system}
First, however, one has to realize that a measurement of the field 
components in this curved metric only makes sense in a local 
Cartesian coordinate frame ({\sc onf}) on the world line of an 
observer, as was discussed in Sec. \ref{sectedrad}. 
It is therefore customary to employ a non-coordinate frame for an observer at rest with respect to the {\sc gw}, 
the tedrad $\lambda_{(\alpha)}^{\mu}$, (\cite{ciufolini}, \cite{weinberg}, \cite{kibble} and \cite{stefani}). 
For these tedrad vectors to 
form a orthonormal frame, they have to obey the orthonormality 
condition:
\bq\label{loodrecht}
g_{\mu\nu}\lambda_{(\alpha)}^{\mu}\lambda_{(\beta)}^{\nu}=\eta_{(\alpha\beta)}
\eq 
In physical terms, this is nothing more than a local coordinate transformation to an 
inertial system. 

In this case, the first components of $\lambda^{\mu}_{(2)}$ and 
$\lambda^{\mu}_{(3)}$ can be chosen to be zero, and the last 
component of $\lambda^{\mu}_{(2)}$ to be zero to remove the 
arbitrarity of a rotation around the $x$-axis. The other components 
then follow from (\ref{loodrecht}). 

\bqa
\lambda_{(0)}^{\mu} = (1+\half h_{00}, 0, 0, 0) &,& \lambda_{(0)\mu} = (\half h_{00} -1, h_{10}, h_{20}, h_{30})  
\\\nonumber
\lambda_{(1)}^{\mu} = (h_{10}, 1-\half h_{11}, -h_{12}, -h_{13}) &,& \lambda_{(1)\mu} = (0, 1+ \half h_{11}, 0, 0)\\\nonumber
\lambda_{(2)}^{\mu} = (h_{20}, 0, 1-\half h_{22}, 0) &,& \lambda_{(2)\mu} = (0, h_{12}, 1+\half h_{22}, h_{23})\\\nonumber
\lambda_{(3)}^{\mu} = (h_{30}, 0, -h_{23}, 1-\half h_{33}) &,& \lambda_{(3)\mu} = (0, h_{13}, 0, 1+\half h_{33})
\eqa 
The decomposition of the electromagnetic tensor in this {\sc onf} is:


\bq
F_{(\alpha\beta)} = F^{\mu\nu}\lambda_{(\alpha)\mu}\lambda_{(\beta)\nu}
\eq
which for the electric field components results in:


\bqa
E_x &=& E_x^{(0)} - B_{y}^{(0)} h_{30} + B_{x}^{(0)} h_{20} \\\nonumber
E_y &=& E_y^{(0)} \haak{1-\half(h_{00} - h_{22})} \\\nonumber
&-& \haak{\frac{\alpha x}{2ik} - 
\frac{\alpha}{2k^2}}\mbox{e}^{ik(x-ct)} - 
 A\mbox{e}^{ik(x-ct)} + B\mbox{e}^{-ik(x+ct)} \\\nonumber
&+& E_z^{(0)} h_{23} + B_{x}^{(0)} h_{30} - B_{z}^{(0)} h_{10} + 
E_x^{(0)} h_{12}\\\nonumber
E_z &=& E_z^{(0)} \haak{1-\half(h_{00} - h_{33})} \\\nonumber 
&-& \haak{\frac{\beta x}{2ik} - 
\frac{\beta}{2k^2}}\mbox{e}^{ik(x-ct)} 
- C\mbox{e}^{ik(x-ct)} + D\mbox{e}^{-ik(x+ct)} \\\nonumber
&+& E_x^{(0)} h_{13} + B_{y}^{(0)} h_{10} - B_{x}^{(0)} h_{20}
\eqa
and for the magnetic field in:

\bqa
B_{x} &=& B_{x}^{(0)} + E_z^{(0)} h_{12} - E_y^{(0)} h_{13} \\\nonumber
B_{y} &=& B_{y}^{(0)} \haak{1+\half(h_{33} + h_{11})} \\\nonumber
&+& \frac{\beta x}{2ik}\mbox{e}^{ik(x-ct)}+ C\mbox{e}^{ik(x-ct)} + 
D\mbox{e}^{-ik(x+ct)} \\\nonumber
B_{z} &=& B_{z}^{(0)} \haak{1+\half(h_{22} + h_{11})} \\\nonumber
&-& \frac{\alpha x}{2ik}\mbox{e}^{ik(x-ct)}- A\mbox{e}^{ik(x-ct)} - 
B\mbox{e}^{-ik(x+ct)} - B_{y}^{(0)} h_{23} 
\eqa

\newpage
\subsubsection{Wave component solutions}
Now that the electromagnetic field components have been put in a 
non-coordinate basis,  
the previously mentioned continuity conditions are applied.
The values for the arbitrary constants can be computed by requiring 
that in regions $I$ and $III$ only outgoing waves exist ($A_I = C_I = 
B_{III} = D_{III} =0$). This is done in Appendix \ref{constants}.
The solutions of the electromagnetic waves in terms of these 
constants are given below for the three regions under consideration 
(see Figure \ref{fig:2}):

\paragraph{Region I}


\bqa
E_y &=& - B_{z} = \half \haak{\mbox{e}^{-2ikL} -1}\mbox{e}^{-ik(x+ct)} 
\\\nonumber
&\times& \krul{\frac{\alpha}{2k^2} + 
\half(B_{z}^{(0)} - E_y^{(0)})(\xi_{00} - \xi_{22}) 
+ (E_z^{(0)} + B_{y}^{(0)})\xi_{23}
+ (B_{x}^{(0)}\xi_{30} + E_x^{(0)}\xi_{12})} \\\nonumber
E_z &=& + B_{y} = \half \haak{\mbox{e}^{-2ikL} -1}\mbox{e}^{-ik(x+ct)} 
\\\nonumber
&\times & \krul{\frac{\beta}{2k^2} - 
\half(B_{y}^{(0)} + E_z^{(0)})(\xi_{00} - \xi_{33}) 
- (B_{x}^{(0)}\xi_{20} - E_x^{(0)}\xi_{13})} 
\eqa

\paragraph{Region II}


\bqa
E_y &=& E_y^{(0)} -\frac{\alpha(x+L)}{2ik}\mbox{e}^{ik(x-ct)} +  
\mbox{e}^{i(\half \pi - kct)} \sin{kx} \\\nonumber
&\times & \krul{\frac{\alpha}{2k^2} + 
\half(B_{z}^{(0)} - E_y^{(0)})(\xi_{00} - \xi_{22}) 
+ (E_z^{(0)} + B_{y}^{(0)})\xi_{23}
+ (B_{x}^{(0)}\xi_{30} + E_x^{(0)}\xi_{12})} \\\nonumber
B_{z} &=& B_{z}^{(0)} -\frac{\alpha(x+L)}{2ik}\mbox{e}^{ik(x-ct)} -  
\mbox{e}^{i(\half \pi - kct)} \sin{kx} \\\nonumber
&\times & \krul{\frac{\alpha}{2k^2} + 
\half(B_{z}^{(0)} - E_y^{(0)})(\xi_{00} - \xi_{22}) 
+ (E_z^{(0)} + B_{y}^{(0)})\xi_{23}
+ (B_{x}^{(0)}\xi_{30} + E_x^{(0)}\xi_{12})} \\\nonumber
E_z &=& E_z^{(0)} -\frac{\beta(x+L)}{2ik}\mbox{e}^{ik(x-ct)} +  
\mbox{e}^{i(\half \pi - kct)}\sin{kx}  \\\nonumber
&\times & \krul{\frac{\beta}{2k^2} - 
\half(E_z^{(0)} + B_{y}^{(0)})(\xi_{00} - \xi_{33}) 
- (B_{x}^{(0)}\xi_{20} - E_x^{(0)}\xi_{13})} \\\nonumber
B_{y} &=& B_{y}^{(0)} + \frac{\beta(x+L)}{2ik}\mbox{e}^{ik(x-ct)} +  
\mbox{e}^{i(\half \pi - kct)}\sin{kx}  \\\nonumber
&\times & \krul{\frac{\beta}{2k^2} - 
\half(E_z^{(0)} + B_{y}^{(0)})(\xi_{00} - \xi_{33}) 
- (B_{x}^{(0)}\xi_{20} - E_x^{(0)}\xi_{13})}
\eqa

\paragraph{Region III}


\bqa
E_y &=& B_{z} = - \frac{\alpha L}{2ik} \mbox{e}^{ik(x-ct)} \\\nonumber
E_z &=& - B_{y} = - \frac{\beta L}{2ik} \mbox{e}^{ik(x-ct)} 
\eqa
If the second part of the device is set up the same way as before, 
such that the length of this part is a half integer multiple of the 
wavelength ($L=n\pi/k=\half n\lambda$), the retreating 
electromagnetic waves vanish, and what remains are just two linearly 
polarized electromagnetic waves propagating in the positive 
$x$-direction. This is just what one expects, as it is the reverse 
process of the {\sc emw} to {\sc gw} conversion in the first part of 
the detector (in accordance to the remark above (\ref{tem})).


\subsection{Efficiency}\label{vacend}

It is possible to set up an arrangement where one generates an 
electromagnetic wave, which produces a gravitational wave. The 
remaining light is then screened off, while the gravitational wave is 
reconverted into an electromagnetic wave. A measurement of the thus 
produced electromagnetic intensity would proof the intermediate 
existence of a gravitational wave. This intensity, in the 
magnetostatic case, is given by:


\bqa\label{effiexact}
W &=& \frac{A^2}{4\pi} \haak{\frac{G L L^{\prime}}{c^4}}^2 
\brak{(B_{z} B_{z}^{(0)} + B_{y} B_{y}^{(0)})^2 + (B_{y} B_{z}^{(0)} - B_{z} 
B_{y}^{(0)})^2} \\\nonumber 
&=& \frac{A^2}{4\pi} 
\haak{\frac{G}{c^4}}^2 (L B_{z})^4
\eqa

where in the last line the length and the electromagnetic field of 
the two regions are taken to be equal. One can see that the final 
intensity depends on the fourth power of both the length and the 
magnetic field used (and on the square of the intensity of the 
initial electromagnetic field $A$). Still, the smallness of the 
factor $(G/c^4)^2$ does not make this a very feasible terrestial 
detection device at present. 

However, it will be shown in the next few sections, that the astrophysical relevance of 
the discussed {\sc gw} to {\sc emw} conversion is significant. In most cases, {\sc gw} to {\sc emw} conversions will occur (and not the reverse process) the efficiency of which is given by the square root of (\ref{effiexact}). For a magnetostatic background field, this, of course, results in the same efficiency as in Sec. \ref{gerst} and Sec. \ref{label1}. 

The main purpose of the calculations in this section was to rigorously derive 
the characteristics of {\sc gw} $\Leftrightarrow $ {\sc emw} 
conversions in the most general circumstances.

\newpage

\section{Magnetars and binary mergers}\label{magnetar}

In the previous sections, the possibility of converting gravitational 
wave energy into light in the presence of a static, homogeneous 
electromagnetic field was discussed. Such a conversion is interesting 
for two reasons. First, it could be relevant for the energy excess in the 
electromagnetic spectrum of very energetic phenomena such as gamma 
ray bursts. The reason for this is that a sub-class of {\sc grb}s is 
supposed to be powered by binary mergers, which release most of their 
energy in {\sc gw} and not in {\sc em} radiation. 

Secondly, it could be a useful indirect detection 
device of gravitational waves, because {\sc emw}s are much easier to 
detect then {\sc gw}s.

One should therefore look for astrophysical sources with the 
following ingredients:
\btel
\item Obviously, one would like to have a source that produces 
gravitational waves that carry large amounts of energy, so that even 
if only a small fraction is transferred to {\sc emw}s, the result is 
significant, 
\item The source has to produce {\sc gw}s with relatively high 
frequencies ($\approx 10$\ kHz). Otherwise, the {\sc emw}s have a 
frequency (the same as the {\sc gw}) below the interstellar plasma 
frequency and will be absorbed before reaching earth,
\item The interaction will only take place in either an extremely 
strong magnetic field, or a (weaker) field that prevails over 
extremely large distances (i.e. a primordial field).
\item The region of interaction is a vacuum or a thin plasma, so that 
one can neglect the difference of the dispersion relation for the 
{\sc emw} from vacuum and {\sc gw}-dissipations. 
\etel

The most promising candidate, satisfying these demands, is the so 
called magnetar, a compact neutron 
star formed by an a-symmetic supernova collapse, a merger or collision 
of two neutron stars as studied by Kokkotas et. al. 
(\cite{kokkotas92}-\cite{kokkotas99}). 

What remains after such a collapse, is a violently oscillating 
neutron star, quaking in several distinct quasi-normal modes. 
Post-Newtonian fluid modes, such as the $f$-mode (fundamental mode), 
$p$-mode (pressure as restoring force) and $g$-modes (gravity as 
restoring force), are dominated by the fundamental mode with typical 
frequencies of $1-2$Hz.
The socalled $w$ave modes, on the other hand, are purely general 
relativistic modes. These are not fluid modes, but 
{\sc gw}s due to strong space-time dynamics, with typical 
frequencies of $8-12$kHz (and up to $38$kHz). This means that the $w$ 
modes might be able to produce {\sc emw}s in the right frequency 
domain (even without photon acceleration).

The energy radiated by these {\sc gw}s is a fraction of a solar rest 
mass energy and is radiated in a small damping time in axisymmetric 
waves, producing a very large gravitational flux density.
 
Finally, the typical magnetic fields of such magnetars are the 
largest ever encountered: $\approx 10^{16}$ Gauss. If one now assumes 
that the far field of the magnetar can be regarded as a vacuum (Sec. 
\ref{effi}) or a thin plasma (Sec. \ref{plasma}), all our conditions 
are satisfied. 

The next subsection gives an estimate of the produced electromagnetic 
energy flux due to gravitational waves in the presence of a magnetar 
using the results of Sec. \ref{gerst} and Sec. \ref{gravphot}. Then 
in Sec. \ref{plasma} the same will be done for a thin plasma using a 
slightly different approach in which the Maxwell equations are 
already put in a non-coordinate basis before solving them. This 
allows easy comparison of the current densities in the plasma with 
the gravitational effects induced by the {\sc gw}.

\newpage
\subsection{Estimate of efficiency for magnetar}\label{effi}

\begin{figure}[h!]
\begin{center}\includegraphics[width=12cm]{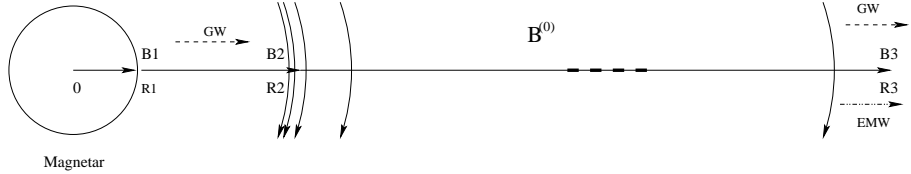}
\caption{\label{fig:magnetar} Gravitons to photons around magnetar.}
\end{center}
\end{figure}

In Figure \ref{fig:magnetar}, the (far) magnetic field of a magnetar 
with mass $M=M_{\odot}$, radius $R_{1}\approx R_{S} = 2GM/c^2 =3\ $km 
and surface field $B_{1}=10^{16}$Gauss is shown. 

The magnetic field 
of such a neutron star will be that of a dipole out to the light 
cylinder and a spherical $1/r$ decay beyond that. Since, however, the 
studies \cite{kokkotas92}-\cite{kokkotas99} involve a non-rotating 
star, only the dipole field will be of importance. 

Furthermore, one can only speak of a {\sc gw} in the radiation zone, for $r \gg R_1$.
In this estimate, following Brodin, Marklund et. al. \cite{brodin00}, it will be assumed that the interaction becomes 
effective around $R_{2}=60R_{S}=180\mbox{km}$.	
 
\bqa
B(r) &=& B_{1}\haak{\frac{R_{1}}{r}}^3 \mdoos{,}  R_{1} < r < R_{3} 
\eqa
To compute the amplitudes of the generated {\sc emw}s, use 
(\ref{gersteffi}) or (\ref{effi2}). To 
apply this formula to the reverse case, realize that the efficiency 
of {\sc gw} to {\sc emw} conversion in a static magnetic field is the 
same as for the reverse case, since the relation connecting 
$\phi_{\mu\nu} \sim h_{\mu\nu}$ and $F_{\mu\nu}$ is linear and symmetric under time reversal. So, for 
$c=G=1$,  $b=\bar{E}=-\bar{B}$, and $a(z)/k = \bar{h}$, where 
$\bar{h}$ is the dimensionless effective amplitude of a $+$-polarised 
plane {\sc gw}, the efficiency is:


\bqa\label{vacgw}
\sqrt{\alpha^{\prime}} &=& \mid\frac{\bar{E}}{\bar{h}}\mid  = 
\frac{-ik}{2}\int \hat{B}(z) dz 
\eqa
And using the {\sc gw} amplitude as calculated by Kokkotas et. al. in \cite{kokkotas96}:
\bqa
\bar{h} &=& 10^{-21} \haak{\frac{E}{10^{-6}M_{\odot}c^2}}^{1/2} 
\haak{\frac{10\mbox{kHz}}{f}}^{1/2} 
\haak{\frac{50\mbox{kpc}}{r}}
\eqa
the {\sc emw} amplitudes can be calculated:
\bqa
E_y (r, t) &=& - B_x (r, t) \\\nonumber
&=& 
-\frac{ik\bar{h}B_1}{2} \haak{\int_{R_2}^{r} 
\haak{\frac{R_{1}}{r^{\prime}}}^3 dr^{\prime}} 
\mbox{e}^{ik(r-t)} \\\nonumber
&=& ik\bar{h}B_1 R_1^3 \haak{\frac{1}{r^2} - \frac{1}{R_2^2}} 
\mbox{e}^{ik(r-t)} 
\eqa
The most energetic magnetars, with $E=10^{-2}M_{\odot}c^2$ will have 
incident {\sc gw}s at $r=180$km with an effective amplitude of $\bar{h}=0.001$. 
Putting this into the equation for the produced electromagnetic waves 
and choosing some cut-off distance $R_3$ where the magnetic field, 
or just the coupling, becomes negligible (or $1/R_3^2$ is just 
negligible with respect to $1/R_2^2$), the result is:


\bq
E_{y\mbox{\tiny max}} = - B_{x\mbox{\tiny max}} \sim 10^{12}\mbox{V/m}
\eq
using $k=\omega$ with $\omega/2\pi=10$kHz. 

This large value results mainly because of the extremely large 
magnetic field, which is four orders of magnitude larger than for 
instance the field of a neutron star binary. 

\subsection{Comparison with neutron star binary} 
If, in (\ref{vacgw}), a close binary just before merger is considered 
as a source, with the same distance scales, but a magnetic surface 
field of $10^{12}$Gauss, an effective (Newtonian) {\sc gw} amplitude 
of $4 \cdot 10^{-4}$ and a {\sc gw} frequency of $2\cdot 10^3$rad/s 
(caused by the rotation of the binary instead of the magnetar 
star-quakes), the {\sc emw} amplitude is still very significant with 
$50$MV/m. 

The frequency of the latter {\sc emw}, however, causes these waves to 
be damped by the interstellar plasma, whereas the former could easily 
be detected by a space based detector (avoiding the ionospheric 
cutoff), as long as the source resides close to our galaxy. In fact, 
for the extreme case discussed here (maximum magnetic field and 
maximum gravitational wave energy), the {\sc emw} amplitude reaching 
earth from a source at a distance of $\approx 50$kpc would be as much 
as $1$V/m if the {\sc emw} only decayed by spherical attenuation.

It has to be realized, though, that the values mentioned here are 
probably highly exaggerated when one takes into account 
all the real dynamics around a magnetar or neutronstar binary. 
More research has to be done to deal with spherically decaying 
{\sc gw}s, non-uniform plasmas etc.

\subsection{Improvements to calculation}
As a first step towards a more realistic astrophysical setting, the 
vacuum has to be replaced by a thin plasma.
In the next sections (\ref{seconf}--\ref{wave}), it 
becomes apparent, that the interaction is hardly effected by the 
presence of such a plasma. The presence of a plasma induces a slight wavenumber offset as compared to 
the vacuum waves, but as long as the coherence lengths of 
the {\sc gw} and {\sc emw} are longer than the extension of the static magnetic field, 
the equations for $E_y (r, t) = - B_x (r, t)$ will remain the same as 
for the vacuum case. The main importance of adding a plasma is to either damp or
excite the {\sc emw}s to higher frequencies.
 
In Sec. \ref{kuijp}, the plasma will be considered in the 
magnetohydrodynamic approximation of a perfectly conducting ideal fluid with pressure gradients. A 
dispersion relation is derived in Sec. \ref{dispersion}, in search of 
plasma waves such as Alfv\'en waves, magnetosonic waves and slow and 
fast modes, excited by the {\sc gw}s.

\newpage
\section{Plasma calculations}\label{plasma}


To obtain the expressions for the {\sc emw} produced by a {\sc gw} 
traveling through a plasma in more or less the same fashion as in 
Section \ref{gwemw}, the Maxwell equations are written in terms of a 
general local orthonormal frame. This will be done in the next 
section \ref{seconf}. In Sec. \ref{currents} the {\sc onf} will 
be specified for a plane polarized {\sc gw}.  

\subsection{Non-coordinate Maxwell equations}\label{seconf}

The general, coordinate, decomposition of the Faraday tensor with 
respect to an observer moving with a velocity $u^a$, satisfying $u_a 
u^a =-1$ is:\footnote{From now $c=1$, roman indices are $0, 1, 2, 3$ and greek indices $1, 2, 3$ following the conventions in \cite{brodin00}.}

\bqa
F^{ab} &=& u^a E^b - u^b E^a + \epsilon^{abc} B_c \mdoosje{or} 
\\\nonumber
E_a &=& F_{0a} \mdoosje{,} B_a = \half\epsilon_{abc} F^{bc} 
\eqa
where $\epsilon_{abc} = u^d \eta_{abcd}$ is the volume element on 
hypersurfaces orthogonal to $u^a$, which is skew symmetric in all its 
indices and satisfies $\epsilon_{123}=\eta_{0123}=\sqrt{|g|}=1$.

To evaluate the Maxwell equations in a locally inertial non-coordinate 
system, again an orthonormal tedrad is used: ${\bf e}_0 = u$ and 
${\bf e}_{\alpha} = \nabla_{\alpha}$ (with $e_a^{\ i} e_b^{\ j} 
g_{ij} = \eta_{ab}$). 
In this {\sc onf} the covariant Faraday tensor is just:

\bq
F_{ab} = u_a E_b - u_b E_a + \epsilon_{abc} B^c
\eq
and the covariant derivatives are defined, completely analogous to 
the usual tensor form in terms of Christoffel symbols, as:

\bqa
\nabla_a F_{bc} &=& \ted_a (F_{bc}) - \ric{d}{ba} F_{dc} - 
\ric{d}{ca} F_{bd} \\\nonumber
\nabla_a F^{bc} &=& \ted_a (F^{bc}) + \ric{b}{da} F^{dc} + 
\ric{c}{da} F^{bd} \\\nonumber
\eqa
where in the non-coordinate desciption, the Christoffel symbols are 
replaced by the socalled {\em Ricci rotation coefficients}. These are 
defined by:

\bq
\ric{c}{ab} = e^c_{\ i} e_b^{\ j} \nabla_j e_a^{\ i} 
\mdoos{$\Leftrightarrow $} \nabla_{\ted_b} \ted_a = \ric{c}{ab} 
\ted_c \mdoos{and} \Gamma_{(ab)c}=0
\eq
The general form of the inhomogeneous Maxwell equations in the {\sc 
onf} can now be expressed as:

\bqa
\nabla_a F^{ba} &=&
\ted_a (F^{ba}) + \ric{b}{da} F^{da} + \ric{a}{da} F^{bd} \\\nonumber
&=& \ted_a (u^b E^a) - \ted_a (u^a E^b) + \epsilon^{bac} \ted_a (B_c) 
\\\nonumber
&+& \ric{b}{da} u^d E^a - \ric{b}{da} u^a E^d + 
\epsilon^{dac}\ric{b}{da} B_c \\\nonumber
&+& \ric{a}{da} u^b E^d - \ric{a}{da} u^d E^b + \epsilon^{bdg} 
\ric{a}{da} B_g \\\nonumber
&=& 4\pi (j_m)^b
\eqa
and similarly for the homogeneous equations. 
In evaluating these expressions, one has to realize that:
\btel
\item $E^a = E^{\alpha}$ and $B^a = B^{\alpha}$, so all terms in 
$E^0$ or $B^0$ vanish,
\item the time axis is chosen such that $\ted_0 = u$, so the observer 
is at rest with respect to the {\sc onf} and only $u^0=1$ is 
nonvanishing (all terms $u^{\alpha}$ are zero),
\item $\epsilon^{0ab}=u_d\eta^{0abd}=u_0 \eta^{0ab0}=0$ and similarly 
$\epsilon^{a0b}=\epsilon^{ab0}=0$,
\item since the rotation coefficients are skew symmetric in the first 
two indices $\Gamma_{aab}=0$.
\etel

\paragraph{Inhomogeneous equations}
Using these `selection rules', one finds for the first inhomogeneous 
Maxwell equation ($\nabla_a F^{0a}=4\pi\rho$):

\bqa\label{maxre}
\nabla_a F^{0a} &=& \ted_a(E^a) - \ric{0}{\delta 0} E^{\delta}  
+\ric{0}{\delta 0}E^{\delta} + \ric{\alpha}{\delta\alpha}E^{\delta} 
+ \epsilon^{\delta\alpha\gamma}\ric{0}{\delta\alpha}B_{\gamma} \\\nonumber
4\pi \rho_m
\\\nonumber
\nabla \cdot {\bf E} &=& (-\ric{\alpha}{\beta\alpha}E^{\beta} - 
\epsilon^{\alpha\beta\gamma}\ric{0}{\alpha\beta}B_{\gamma}) + 
4 \pi \rho_m \\\nonumber
&=& \rho_E + 4\pi\rho_m
\eqa
where the last equation {\em defines} the extra `charge density' $\rho_E$.
The other three inhomogeneous equations follow from $\nabla_a 
F^{\beta a} = 4\pi j^{\beta}$:

\bqa\label{maxje}
\nabla_a F^{\beta a}  &=&
-\ted_0(E^{\beta}) + 
\epsilon^{\beta\alpha\gamma}\ted_{\alpha}(B_{\gamma}) \\\nonumber
&+& (\ric{\beta}{0\alpha} E^{\alpha} - \ric{\beta}{\delta 
0}E^{\delta}) - \ric{\alpha}{0 \alpha}E^{\beta} \\\nonumber
&+& (\epsilon^{\delta\alpha\gamma}\ric{\beta}{\delta\alpha} + 
\epsilon^{\beta\delta\gamma}\ric{\alpha}{\delta\alpha})B_{\gamma} + 
\epsilon^{\beta\delta\gamma}\ric{0}{\delta 0} B_{\gamma} \\\nonumber
\ted_0({\bf E}) - \nabla \times {\bf B} &=&
(\ric{\alpha}{0\beta}-\ric{\alpha}{\beta 0})E^{\beta} -
\ric{\beta}{0\beta} E^{\alpha} \\\nonumber
&+& \epsilon^{\alpha\beta\gamma}(\ric{0}{\beta 0}B_{\gamma} + 
\ric{\delta}{\beta\gamma}B_{\delta}) -4\pi {\bf j}_m \\\nonumber
&=&-{\bf j_E} - 4\pi {\bf j}_m
\eqa
where the intermediate result that
$(\epsilon^{\delta\alpha\gamma}\ric{\beta}{\delta\alpha} + 
\epsilon^{\beta\delta\gamma}\ric{\alpha}{\delta\alpha})B_{\gamma}  =
\epsilon^{\beta\delta\gamma}\ric{\alpha}{\delta\gamma}B_{\alpha}$ is 
proved in Appendix \ref{symmetric}, and where the last equation defines ${\bf j}_E$.
\newpage
\paragraph{Homogeneous equations}
The first homogeneous Maxwell equation follows from chosing 
$(abc)=(\alpha\beta\gamma)$:

\bqa\label{maxrb}
\nabla_{(\alpha}F_{\beta\gamma)} &=& 
\nabla_{\alpha} F_{\beta\gamma} + \nabla_{\gamma}F_{\alpha\beta} + 
\nabla_{\beta}F_{\gamma\alpha} \\\nonumber
&=& \epsilon_{\beta\gamma\delta}\ted_{\alpha}(B^{\delta}) + 
\epsilon_{\alpha\beta\delta}\ted_{\gamma}(B^{\delta}) + 
\epsilon_{\gamma\alpha\delta}\ted_{\beta}(B^{\delta}) 
\\\nonumber
&-& \ric{d}{\beta\alpha}F_{d\gamma} -\ric{d}{\alpha\gamma}F_{d\beta} 
- \ric{d}{\gamma\beta}F_{d\alpha} \\\nonumber
&-& \ric{d}{\gamma\alpha}F_{\beta d} - \ric{d}{\beta\gamma}F_{\alpha 
d} - \ric{d}{\alpha\beta}F_{\gamma d} \\\nonumber
-\ted_{\delta}(B^{\delta})&=& 
   \ric{0}{\alpha\beta}E_{\gamma} + \ric{0}{\gamma\alpha}E_{\beta} + 
\ric{0}{\beta\gamma}E_{\alpha} \\\nonumber
&-&\ric{0}{\beta\alpha}E_{\gamma} - \ric{0}{\alpha\gamma}E_{\beta} - 
\ric{0}{\gamma\beta}E_{\alpha} \\\nonumber
&+& \epsilon_{\gamma \delta\nu}\ric{\delta}{\beta\alpha}B^{\nu} 
+   \epsilon_{\beta \delta\nu}\ric{\delta}{\alpha\gamma} B^{\nu}
+   \epsilon_{\alpha \delta\nu}\ric{\delta}{\gamma\beta} 
B^{\nu}\\\nonumber
&-& \epsilon_{\alpha \delta\nu}\ric{\delta}{\beta\gamma} B^{\nu}
-   \epsilon_{\gamma \delta\nu}\ric{\delta}{\alpha\beta} B^{\nu}
-   \epsilon_{\beta \delta\nu}\ric{\delta}{\gamma\alpha} 
B^{\nu}\\\nonumber
&=& \epsilon^{\alpha\beta\gamma}\ric{0}{\alpha\beta}E_{\gamma} -     
\epsilon^{\alpha\beta\gamma}\epsilon_{\gamma\delta\nu}\ric{\delta}{\alpha\beta}B^{\nu}\\\nonumber
&=& \epsilon^{\alpha\beta\gamma}\ric{0}{\alpha\beta}E_{\gamma} + 
    \ric{\beta}{\nu\beta}B^{\nu} 
\mdoos{($+\ric{\alpha}{\alpha\beta}B^{\beta}=0$)} \\\nonumber
\nabla \cdot {\bf B} &=& -\ric{\alpha}{\beta\alpha}B^{\beta} - 
\epsilon^{\alpha\beta\gamma}\ric{0}{\alpha\beta}E_{\gamma} = \rho_B
\eqa
The other homogeneous Maxwell equations follow from chosing 
$(abc)=(0\alpha\beta)$:

\bqa\label{maxjb}
\nabla_{(0} F_{\alpha\beta )} &=&
\nabla_0 F_{\alpha\beta} + \nabla_{\beta}F_{0\alpha} + 
\nabla_{\alpha}F_{\beta 0} \\\nonumber
&=& (\ted_{\beta}(E_{\alpha}) - \ted_{\alpha}(E_{\beta})) + 
\epsilon_{\alpha\beta\gamma}\ted_0(B^{\gamma}) \\\nonumber
&-& (\ric{d}{\alpha 0}F_{d\beta} + \ric{d}{\beta 0}F_{\alpha d}) 
\\\nonumber
&-& (\ric{d}{0\beta}F_{d\alpha}+\ric{d}{\alpha\beta}F_{0d}) - 
(\ric{d}{\beta\alpha}F_{d0} + \ric{d}{0\alpha}F_{\beta d}) \\\nonumber
-\ted_0({\bf B})+\nabla\times{\bf E}  &=&
-(\ric{0}{\alpha 0}E_{\beta} - \ric{0}{\beta 0}E_{\alpha}) - 
\ric{0}{0\beta}E_{\alpha} \\\nonumber
&-&(\ric{\delta}{\alpha\beta}E_{\delta} 
-\ric{\delta}{\beta\alpha}E_{\delta}) - \ric{0}{0\alpha}E_{\beta} 
\\\nonumber
&+& (\epsilon_{\beta\delta\gamma}\ric{\delta}{\alpha 0} - 
\epsilon_{\alpha\delta\gamma}\ric{\delta}{\beta 0})B^{\gamma} 
\\\nonumber
&+& (\epsilon_{\alpha\delta\gamma}\ric{\delta}{0 \beta} - 
\epsilon_{\beta\delta\gamma}\ric{\delta}{0\alpha})B^{\gamma} 
\\\nonumber
&=& -\epsilon^{\alpha\beta\gamma}(\ric{0}{\beta 0}E_{\gamma} + 
\ric{\delta}{\beta\gamma}E_{\delta}) \\\nonumber
&+& 
(\epsilon^{\nu\beta\alpha}\epsilon_{\alpha\delta\gamma}\ric{\delta}{0\beta} 
- 
\epsilon^{\nu\beta\alpha}\epsilon_{\alpha\delta\gamma}\ric{\delta}{0\beta})B^{\gamma} 
\\\nonumber
\ted_0({\bf B})-\nabla\times{\bf E}&=& 
\epsilon^{\alpha\beta\gamma}(\ric{0}{\beta 0}E_{\gamma} + 
\ric{\delta}{\beta\gamma}E_{\delta}) \\\nonumber
&+& (\ric{\alpha}{0\beta} - \ric{\alpha}{\beta 0})B^{\beta} - 
\ric{\beta}{0\beta}B^{\alpha} \\\nonumber
&=& -{\bf j}_B
\eqa

\newpage

\subsection{Gravity induced charge and current 
densities}\label{currents}

In this section, the {\sc onf} used in the non-coordinate Maxwell 
equations (\ref{maxre})-(\ref{maxjb}), is specified to a plane, 
$+$-polarized {\sc gw} with metric:
diag$(-1, (1+h), (1-h), 1)$ (compare with (\ref{tt})). The natural 
tedrad for this system is, as one can easily verify by applying the 
orthonormality condition (\ref{loodrecht}):

\bqa\label{gwmetric}
e_{(0)}^{\quad i} &=& (1, 0, 0, 0)) \\\nonumber
e_{(1)}^{\quad i} &=& (0, (1-\half h), 0, 0) \\\nonumber
e_{(2)}^{\quad i} &=& (0, 0, (1+\half h), 0) \\\nonumber
e_{(3)}^{\quad i} &=& (0, 0, 0, 1) 
\eqa
In this {\sc onf} the only nonzero rotation coeffiecients are 
$\ric{1}{13}=-\ric{2}{23}=\half \frac{\partial h}{\partial z}$ and 
$\ric{1}{10}=-\ric{2}{20}=-\half \afg{h}$, since only the $z$ and $t$ 
derivatives of the $\ted_2$ and $\ted_3$ vectors contribute. 

This imediately tells one that the gravity induced charge densities 
vanish. The only components that remain for the induced current 
densities are:
\bqa
{\bf j}_E &=& \left[ \ric{\alpha}{\beta 0}E^{\beta} - 
\epsilon^{\alpha\beta\gamma}\ric{\delta}{\beta\gamma}B_{\delta}\right] 
\ted_{\alpha} \\\nonumber
{\bf j}_B &=& \left[ \ric{\alpha}{\beta 0}B^{\beta} - 
coo\epsilon^{\alpha\beta\gamma}\ric{\delta}{\beta\gamma}E_{\delta}\right] 
\ted_{\alpha}
\eqa
Explicitely, the second term in the equation for ${\bf j}_E$ for 
$\alpha = 1$, $\beta = 2$ and $\gamma = 3$ becomes\footnote{using 
$\dot{h} = \afgz{h} = -\afg{h}$.}:
\bqa
(\ric{\delta}{23}-\ric{\delta}{32}) B_{\delta} &=& 
\ric{\delta}{23}B_{\delta}=(0, (1+\half h), 0) 
\frac{\partial}{\partial z} 
\left(\begin{array}{c}
(1+\half h)B_x \\
(1-\half h)B_y \\
B_z \end{array}
\right) \\\nonumber
&=& -\half \dot{h} B_{y}
\eqa
whereas for $\alpha = 2$, $\beta = 3$ and $\gamma = 1$:
\bqa
(\ric{\delta}{31}-\ric{\delta}{13}) B_{\delta} &=& -\ric{\delta}{13} 
B_{\delta} = -
((1-\half h), 0, 0) \frac{\partial}{\partial z} 
\left(\begin{array}{c}
(1+\half h)B_x \\
(1-\half h)B_y \\
B_z \end{array}
\right) \\\nonumber
&=& -\half \dot{h} B_{x}
\eqa
Similarly from the equation for ${\bf j}_{B}$ (replacing ${\bf B} 
\rightarrow -{\bf E}$) it follows that:
\bqa
-\ric{\delta}{23} E_{\delta} &=& \half \dot{h} E_{y} \mdoos{and} 
\\\nonumber
-\ric{\delta}{31} E_{\delta} &=& \half \dot{h} E_{x}
\eqa
and for the other components of ${\bf j}_E$:

\bqa
\ric{1}{\beta 0}E^{\beta} &=& -\half\afg{h}E_x \\\nonumber
\ric{2}{\beta 0}E^{\beta} &=& \half\afg{h}E_y
\eqa
and for ${\bf j}_B$:
\bqa
\ric{1}{\beta 0}B^{\beta} &=& -\half\afg{h}B_x \\\nonumber
\ric{2}{\beta 0}B^{\beta} &=& \half\afg{h}B_y
\eqa
Putting things together:
\bqa\label{stromen}
{\bf j}_E &=&
-\half\dot{h}
\left(\begin{array}{c}
B_y - E_x \\
B_x + E_y \\
0
\end{array}
\right) \\\nonumber
{\bf j}_B &=& \half\dot{h}
\left(\begin{array}{c}
B_x + E_y \\
B_y - E_x \\
0
\end{array}
\right) 
\eqa
The Maxwell equations in the presence of a gravitational wave are summarized below in their final form for 
future reference.



\begin{center}
\fbox{$
\begin{array}{rcll}\label{dive}
 \nabla \cdot {\bf E}  & = &  4\pi \rho_{m}  & \mbox{\sc (poisson)} 
 \\\label{afge} &&& \\
 \afg{\bf E} - \nabla \times {\bf B}  & = &   -4\pi {\bf j}_{m} 
- {\bf j}_{E}  & \mbox{\sc (ampere)} \\\label{divb} &&&\\
\nabla \cdot {\bf B}  & = &  0  & \mbox{\sc (no monopoles)}\\\label{afgb} &&&\\
 \afg{\bf B} + \nabla \times {\bf E} & = &  -{\bf j}_{B}  & \mbox{\sc (faraday)} 
\end{array}
$}
\end{center}
With ${\bf j}_E$ and ${\bf j}_B$ defined in (\ref{stromen}).

\newpage

\subsection{Energy-momentum conservation}\label{conservation}

To evaluate the Maxwell equations, or to find electromagnetic wave 
equations, one still needs
an expression for the matter current. This expression follows from 
the linearized equation of motion. In this case, for a one component 
plasma where the ion velocity is negligible compared to the electron 
velocity, one can use the energy momentum conservation equation (see 
\cite{goldston} or \cite{kampen}). 

The matter part of the stress energy tensor, for a pressure free 
`dust-like' plasma, is (see \cite{weinberg} and \cite{hawkingellis}) $T_{\mbox{\sc \tiny m}}^{ab} = \rho V^a V^b$ 
(where $V^a = (1, {\bf v})$ is the non-relativistic fluid velocity 
for $\gamma \ll 1$). The electromagnetic part is not needed here because of the useful identity $\nabla_b T^{ab}_{\mbox{\sc \tiny em}} = 
-F^{ab}(j_{\mbox{\sc \tiny m}})_b$. Thus the conservation equations 
follow from:

\bq
\nabla_b (m n V^a V^b) - q n (u^a E^b - u^b E^a + 
\epsilon^{abc}B_c)V_b =0 
\eq
or, in a $1+3$ split,
\bqa\label{eqmotion1}
\afg{n} + \nabla \cdot (n{\bf v}) &=& 
-n(\ric{\alpha}{0\alpha} + \ric{\alpha}{00}v_{\alpha} + 
\ric{\alpha}{\beta\alpha}v^{\beta})  \\\nonumber
\haak{\afg{} + {\bf v}\cdot \nabla }{\bf v} &=& \frac{q}{m} ({\bf E} 
+ {\bf v}\times {\bf B}) 
-(\ric{\alpha}{00} + (\ric{\alpha}{0\beta} + \ric{\alpha}{\beta 
0})v^{\beta} + \ric{\alpha}{\beta\gamma} v^{\beta} v^{\gamma})
\eqa
In the specified metric, these equations can be linearized around the 
unperturbed plasma state: 
${\bf E} = 0$, ${\bf v}^0 =0$ and $\bar{\bf v}(z,t) = \bar{\bf 
v}(z)\mbox{e}^{-i\omega t}$, where all the barred quantities are first order amplitudes. 
In that case, all the terms containing 
the rotation coefficients are either zero or quadratic in $\bar{\bf 
h}$, which reduces the equation of motion (\ref{eqmotion1}) to:

\bq\label{eqmotion1b}
\afg{\bar{\bf v}} = -i\omega \bar{\bf v} = \frac{q}{m} (\bar{\bf E} + 
\bar{\bf v} \times \hat{\bf B}) 
\eq

Now take ${\bf j}_m=q n_0 {\bf v}$ (with $n_0$ the unperturbed 
electron number density) and introduce the following relevant 
frequencies:
\bqa
\mbox{\sc cyclotron frequency:} &\Rightarrow & \omega_c = 
\frac{q\hat{B}}{m} \\\nonumber
\mbox{\sc plasma frequency:} &\Rightarrow & \omega_p = 
\sqrt{\frac{4\pi n_0 q^2}{m}} \\\nonumber
\mbox{\sc upper hybrid frequency:} &\Rightarrow & \omega_h = 
\sqrt{\omega_c^2 + \omega_p^2}
\eqa
and solve (\ref{eqmotion1b}) for $\bar{\bf v}$ or $\bar{\bf 
j}_{\mbox{\sc \tiny m}}$:

\bqa
(\bar{j}_m)_x &=&\frac{1}{4\pi} \frac{\omega_p^2}{\omega}i\bar{E}_x 
\\\nonumber
(\bar{j}_m)_y &=&\frac{1}{4\pi} \frac{\omega_p^2}{\omega} 
\haak{
i\bar{E}_y + \frac{\omega_c}{\omega} \bar{E}_z
}\haak{
\frac{\omega^2}{\omega^2-\omega_c^2}
}\\\nonumber
(\bar{j}_m)_z &=& \frac{1}{4\pi} \frac{\omega_p^2}{\omega} 
\haak{
i\bar{E}_z - \frac{\omega_c}{\omega} \bar{E}_y
}\haak{
\frac{\omega^2}{\omega^2-\omega_c^2}
}
\eqa
The component needed in the next section is: 
\bq\label{jy}
4\pi i\omega (\bar{j}_{\mbox{\sc \tiny m}})_y = - 
\omega_p^2\haak{\frac{\omega^2 - \omega_p^2}{\omega^2 - 
\omega_h^2}}\bar{E}_y
\eq


\subsection{Exact calculation for the plasma waves}\label{wave}

Now that the Maxwell equations  are put in an elegant and transparent 
form, it is straightforward to obtain wave equations for an {\sc emw} excited by a {\sc gw} 
passing through a static magnetic background field ${\bf B}^{(0)}=(\hat{B}(z), 0, 
0)$. For the electric wave components, take the curl of Faraday's law 
(\ref{maxjb}) and substitute it into Amp\`ere's law, differentiated 
with respect to time. And to find magnetic components, proceed the other 
way around. Thus, one finds:

\bqa
\haak{\afgg{} - \nabla^2}{\bf E} &=& - \afg{}(4\pi {\bf j}_m +   
{\bf j}_E) - \nabla\times {\bf j}_B \\\nonumber
\haak{\afgg{} - \nabla^2}{\bf B} &=&  \nabla\times(4\pi {\bf j}_m 
+ {\bf j}_E) - \afg{{\bf j}_B}
\eqa
To solve these equations assume the {\sc gw} to be monochromatic and 
let the frequency of all perturbations coincide with that of the 
driver, leaving the spatial dependence free for the moment, i.e.:
\bqa
{\bf h} &=& \bar{\bf h}\mbox{e}^{ik(z-t)}\\\nonumber 
{\bf E} &=& \bar{\bf E}(z)\mbox{e}^{-i\omega t} \\\nonumber
{\bf j}_m &=& \bar{\bf j}_m(z)\mbox{e}^{-i\omega t}\\\nonumber
{\bf B}&=& \hat{{\bf B}}^{(0)}(z)\ted_1 + 
(\bar{\bf B})^{(1)}(z)\mbox{e}^{-i\omega t}
\eqa
Dropping all second order terms in the perturbation (which are in 
effect just the current density terms in 
$\bar{\bf h}\cdot\bar{\bf E}$), the relevant wave components are:
\bqa\label{ewave}
\haak{\frac{\partial^2}{\partial z^2} + \omega^2}\bar{E}_y + 4\pi i\omega 
(\bar{j}_m)_y &=& 
-\haak{k^2\hat{B}-\half ik\frac{\partial \hat{B}}{\partial 
z}}\bar{h}\mbox{e}^{ikz} \equiv F(z) \\\label{bwave}
\haak{\frac{\partial^2}{\partial z^2} + \omega^2}\bar{B}_x 
+4\pi \frac{\partial}{\partial z}(\bar{j}_m)_y &=& 
\haak{k^2\hat{B}-\half ik\frac{\partial \hat{B}}{\partial 
z}}\bar{h}\mbox{e}^{ikz} = - F(z)
\eqa
where $F(z)$ is a function of $z$ through the specific form of the static 
background field. 

The expression for $\hat{E}_y$ is now easy to evaluate using 
(\ref{jy}), while $\hat{B}_x$ is harder to find, since the 
$z$-dependence of $(j_m)_y$ has not been specified. $\hat{B}_x$ can, 
however, be deduced from $\hat{E}_y$, as will be done below.

\paragraph{Solution for $\hat{E}_y$:}
Inserting (\ref{jy}) in (\ref{ewave}), leads to an inhomogeneous 
differential equation for {\bf E}:
\bq\label{lable}
\haak{\frac{\partial^2}{\partial z^2} + \omega^2 - 
\omega_p^2\haak{\frac{\omega^2 - \omega_p^2}{\omega^2 - 
\omega_h^2}}}\bar{E}_y = F(z)
\eq 
To find the solution of this equation, one first determines the 
solution of the homogeneous equation, taking $F(z)=0$, and its {\em 
Wronskian}:
\bqa\label{dikkiedik}
\bar{E}^{h}_y (z) &=& A\mbox{e}^{ik_l z} + B\mbox{e}^{-ik_l z} 
\\\nonumber
W &=& \mbox{det}\left|
\begin{array}{lr}
A\mbox{e}^{ik_l z} & B\mbox{e}^{-ik_l z} \\
ik_l A\mbox{e}^{ik_l z} & -ik_l B\mbox{e}^{-ik_l z}
\end{array}
\right| = -2ik_l AB 
\eqa
In (\ref{dikkiedik}), $k_l$ is the wavenumber of the {\sc emw} satisfying the 
dispersion equation $k_l^2 = \omega^2 - (\Delta \omega)^2$ which is slightly shifted 
compared to the dispersion relation in vacuum. This shift is given by:
\bq
\Delta \omega = \sqrt{\omega_p^2\haak{\frac{\omega^2 - 
\omega_p^2}{\omega^2 - \omega_h^2}}} \ll \omega
\eq
The general solution of the inhomogeneous equation is then given by:  
\bq\label{dikkie}
\bar{E}_y (z) = \frac{i}{2k_l}\left(
\mbox{e}^{ik_l z} \int_0^z \mbox{e}^{-ik_l z^{\prime}} F(z^{\prime}) 
dz^{\prime} + 
\mbox{e}^{-ik_l z} \int^L_z \mbox{e}^{ik_l z^{\prime}} F(z^{\prime}) 
dz^{\prime}
\right)
\eq
Here, as in Sec. \ref{gerst}, the assumptions have been made that the static background field 
is restricted to a region $0<z<L$ and that there are no incoming {\sc 
emw}'s in the interaction region.

From (\ref{dikkie}), the amplitude of the outgoing wave can be 
calculated by  	
inserting the expression for $F(z)$, expanding terms in $\Delta \omega \ll 
\omega$ and assuming that $k_l$ is approximately constant in the 
interaction region.
\bqa\label{emwe}
(\bar{E}_y)_{\mbox{\tiny out}} &=& -\frac{i}{2k_l}\left\{
\omega^2 \bar{h} - \half i\omega\bar{h} \frac{i\Delta \omega^2}{2\omega}\right\}
\int_0^L \hat{B}(z)\mbox{e}^{\left(\frac{i\Delta \omega^2}{2\omega}\right)z} dz 
\\\nonumber
&\approx& -\frac{i\bar{h}\omega}{2}\frac{\sqrt{k_l}}{k_l}
\int_0^L \hat{B}(z)\mbox{e}^{\left(\frac{i\Delta \omega^2}{2\omega}\right)z} dz 
\\\label{plasmasol}
&\approx& -\frac{i\bar{h}\omega}{2}\int_0^L 
\hat{B}(z)\mbox{e}^{\left(\frac{i\Delta \omega^2}{2\omega}\right)z} dz  
\mdoosje{with} \frac{\sqrt{k_l}}{k_l}\approx 1 
\eqa
where the second term in (\ref{emwe}) follows from partial 
integration of $\afgz{\hat{B}}$ and the stockterm vanishes at the 
boundaries because the magnetic field is localized. The full solution 
for $\bar{E}_y (z,t)$ is then just:

\bq
\bar{E}_y (z,t) = (\bar{E}_y)_{\mbox{\tiny out}}\mbox{e}^{i(k_l z-\omega t)}
\eq

\paragraph{Solution for $\hat{B}_x$:}
The calculation of the magnetic part of the outgoing wave follows 
realizing that all that: 
\bq
4\pi \afgz{}(\bar{j}_m)_y = 4\pi i k_l (\bar{j}_m)_y \approx 4\pi i \omega \haak{1-\half \haak{\frac{\Delta \omega}{\omega}}^2} (\bar{j}_m)_y \approx 4\pi i \omega (\bar{j}_m)_y
\eq
Comparing this with (\ref{ewave}) and {\ref{bwave}), it follows that $\hat{B}_x \approx - \hat{E}_y$, as one expects. 

The {\sc emw}s, or `light', generated by the gravitational waves are apparently hardly effected by the presence of a thin plasma. 
The wavenumber has a small offset with respect to the vacuum solutions, reflecting some dispersion caused by the plasma, but due to the strong magnetic background field, this will be a completely negligible effect. The wavenumber `mismatch' is after all inversely proportional to the cyclotron frequency (in the regime $\omega_p^2/\omega_c \ll \omega \approx \omega_p \ll \omega_c$) and thus also to  the magnetic field, so for a magnetic field of say $10^8$ Gauss and a typical frequency of a few kHz $(\Delta \omega)^2/\omega \sim 10^{-19}$rad/s.

\newpage

\subsection{Magnetars revisited}

As mentioned in the previous section, the presence of a plasma in the 
interaction region is not very important. This is the result of the 
fact that the large magnetic field suppresses the mobility (resulting 
in $\Delta \omega \propto 1/\omega_c \propto 1/\hat{B} \ll 1$). Assuming 
therefore that the coherence length of the {sc gw} and {\sc emw} is larger than the length of the 
interaction region, the same results are found as for the vacuum case.
The difference lies in the fact that in order to travel over 
astrophysical distances, the generated {\sc emw}s have to overcome 
the interstellar plasma. As soon as the magnetic field has decreased 
such that $\omega_c < \omega$, (\ref{eqmotion1b}) and a spherical 
decay of the {\sc emw} beyond the interaction region allow the 
electron quiver velocity to become relativistic at a distance: 

\bqa
v(r) &\sim & \frac{q}{m\omega}\frac{R_3}{r}\bar{E}_{\mbox{\tiny max}} 
\\\nonumber
r_{\mbox{\tiny rel}} &\sim & 3 \cdot 10^{12} \mbox{km}
\eqa
and the electron velocity somewhere in between. In that case the {\sc 
emw}s become highly nonlinear and effects such as parametric 
excitation and, even more importantly, harmonic generation become 
important, which is interesting, because the latter effect can 
convert the {\sc emw}s to even larger frequencies. 

It is not likely that this, non-linear, effect will play a very important r\^ole 
in the interstellar plasma, exciting higher harmonics with $60$ times the frequency of the 
original waves, as has been done in laser experiments in a laboratory plasma. If, however, 
only the first or second higher harmonic is reached (with the same energy), this would be enough 
to overcome the interstellar plasma damping, and the resulting light should be detectable 
with a telescope such as 
the proposed {\em Astronomical Low Frequency Array} ({\sc alfa}).  
Supernov{\ae} and collapsing binaries within the Local Group and the 
Virgo cluster might be detectable in this way with a event rate of as 
many as a few per year as soon as detectors such {\sc alfa} are 
operational. Furthermore, these signals could be compared with the 
{\sc gw} observations when the first {\sc gw} detectors such as {\sc 
ligo, lisa} and {\sc virgo} open the {\sc gw} observation chanel.

\newpage
\section{Magnetohydrodynamics}\label{kuijp}
Another way to study interactions in a plasma is of course in the 
magnetohydrodynamic approximation. Consider an infinitely conductive 
hydrogen-like plasma in a singe fluid approximation (for more 
details, see \cite{goldston} and \cite{kampen}) where a pressure term 
is added in comparison to Sec. \ref{conservation}. In other words, 
consider the full energy-momentum tensor. 

In an incompressible plasma, the unperturbed pressure will be 
negligible with respect to the energy density $\mu=\rho(1+\Pi)$ (with $\rho \Pi$ the, small, internal energy density), but the pressure 
gradients will be important. To avoid mistakes, all terms will be 
maintained throughout the exact calculation.     

\subsection{Conservation equations}
The stress-energy tensor for a plasma in an electromagnetic field can 
naturally be decomposed in the tensor for the energy in the pure electromagnetic field that of the matter field, as in Sec. 
\ref{conservation}:

\bqa
T^{ab} &=& T_m^{ab} + T_{\mbox{\tiny \sc em}}^{ab} \\\nonumber
&=& (\mu + p) V^a V^b + p \eta^{ab} + \frac{1}{4\pi}
\brak{F^{ac}F_c^{\ b} - \frac{1}{4} 
\eta^{ab}F^{cd}F_{cd}} 
\eqa
which makes the conservation equations easy to evaluate (using the 
result of Exercise 3.18 in \cite{gravitation}):


\bqa
\nabla_b T^{ab} &=& \nabla_b T_m^{ab} + \nabla_b T_{\mbox{\tiny \sc 
em}}^{ab} \\
&=& \nabla_b T_m^{ab} - F^{ab} j_b = 0 
\eqa

Resulting, using the $3+1$ split, in the equation of energy 
conservation ($a=0$) and the equations of motion ($a=\alpha= 1, 2, 
3$). Again $V^a=(1, {\bf v})^T$ and $j^a =(\rho_m, {\bf j}_m)^T$.


\bqa
\afg{}(\mu + p) + \nabla\cdot (\mu + p){\bf v} -\afg{p} &=& {\bf j} 
\cdot {\bf E} \\\nonumber
 &-& (\mu + p)\left[\ric{0}{db}V^d V^b + \ric{b}{db}V^d \right] 
\\\nonumber
&-& p \left[\ric{0}{db}\eta^{db} + \ric{b}{db}\eta^{0d} \right] 
\\\label{eomei}
{\bf v} \haak{\afg{}(\mu + p) + \nabla\cdot (\mu + p){\bf v}} &+& 
 (\mu + p)\haak{\afg{} + {\bf v} \cdot \nabla}{\bf v} + \nabla 
p\\\nonumber 
  &=& {\bf j_m} \times {\bf B} - \rho {\bf E} - ({\bf v} \cdot {\bf 
j}_m){\bf E} + ({\bf j}  \cdot {\bf E}){\bf v} \\\nonumber
&-& (\mu + p) \left[ \ric{\alpha}{db}V^d V^b + \ric{b}{db} V^{\alpha} 
V^d \right] \\\nonumber
&-& p \left[\ric{\alpha}{db}\eta^{db} +  \ric{b}{db}\eta^{\alpha 
d}\right] 
\eqa

Most terms in these rather long expressions vanish in a first order 
calculation in the metric of a plane polarised {\sc gw} 
(\ref{gwmetric}):
\begin{itemize}
\item[$\diamond$] All the rotation coefficients in the first equation 
vanish by themselves in this metric,  
\item[$\diamond$] the only non-vanishing Ricci coefficients in the 
second equation result in:
\bq
\half i (\mu + p) (k_g \bar{v}_z + \omega_g) \bar{ h} 
\left(
\begin{array}{r}
\bar{v}_x \\
-\bar{v}_y \\
0
\end{array}
\right)
\eq
which is clearly non-linear in the perturbed quantities,
\item[$\diamond$] both ${\bf E}$ and ${\bf v}$ and ${\bf j}_m$ have 
no zeroth order components in the perturbation, so all the products 
of these quantities are second order or higher, 
\item[$\diamond$] $\rho {\bf E}$ is negligible with respect to ${\bf 
j_m} \times {\bf B}$ due to the charge neutrality in a plasma (the 
global charge density is very small, whereas the currents can be 
substantial; for details, again, see \cite{goldston} or 
\cite{kampen}),
\item[$\diamond$] for the same reason the convective term ${\bf 
v}\cdot\nabla{\bf v}$ vanishes.  
\end{itemize}
A final simplification is made by substituting the first equation in 
the second. What remains are the following simple conservation 
equations:

\bqa\label{eom}
\afg{\bar{\mu}} + (\mu + p)\nabla\cdot\bar{\bf v} &=& 0 
\\\label{energy}
(\mu + p)\afg{\bar{\bf v}} + \nabla\bar{p} &=&\bar{\bf 
j}_m\times\hat{\bf B} \\\label{eomei2}
&=& \frac{1}{4\pi} (\nabla \times \bar{\bf B})\times\hat{\bf B} - 
\frac{1}{4\pi} \bar{\bf j}_E\times\hat{\bf B} 
\eqa 
where in the last line, `Amper\`es' law is inserted without the 
displacement current, which is negligible with respect to the current 
${\bf j}_m \sim \nabla\times{\bf B} \gg \frac{1}{c^2}\afg{\bf E}$ in 
the {\sc mhd} approximation. 
The above equations are just the usual mass conservation equation and 
the equations of motion, but now with added terms that express the 
gravitational effects.

\subsection{Ohms law}
To close the system of equations, one also needs the generalized law 
of Ohm, which follows from the separate equation of motion for the 
electrons in the approximation that the electron inertia is 
negligible ($d{\bf v}_e/dt =0$). This is adequate as long as the 
perturbations caused by the {\sc gw} are slow compared to the motion 
along the magnetic field (a weak constraint for the cyclotron motion 
around an extremely strong magnetic field), so the electrons have 
ample time to reach dynamical equilibrium. For zero resistivity, the 
single-fluid electron equation of motion is then given by (compare to 
(\ref{eomei}) and (\ref{eomei2})): 

\bq
\nabla p_e = \rho_e ({\bf E} + {\bf v}_e \times {\bf B}) 
\eq
or, from ${\bf j}_m \approx \rho_e({\bf v}_i - {\bf v}_e)$ and 
therefore ${\bf v}_e\approx {\bf v} - {\bf v}_i \approx {\bf v} - 
{\bf j}/{\rho_e}$:

\bq
{\bf E} = -{\bf v} \times {\bf B} + \frac{1}{\rho_e} ({\bf j} \times 
{\bf B} - \nabla p_e) 
\eq
However, the more trivial and commonly known, Ohms law ${\bf E} =- 
{\bf v} \times {\bf B}$ prevails in the limit of a small Larmor 
radius with respect to the typical lenght scale of the {\sc gw} 
perturbations (\cite{kampen}). Since the magnetic field is very 
strong, this Larmor radius will be smaller than $10^{-8}$cm for 
electrons moving with thermal velocity ($v_{th, e}\approx 6\cdot 
10^{7}$m/s at $T\sim 10^8$K) and still smaller for lower 
perpendicular velocities. Even the smallest expectable {\sc gw} 
perturbations should therefore be much larger than the Larmor radius.

\subsection{{\sc mhd} Maxwell equations}
The reason for all this is to eliminate the electric field entirely 
from our set of equations, which is achieved by inserting the 
linearized, general law of Ohm in Faraday's equation:   
\bq\label{max}
\afg{\bar{\bf B}} = \nabla\times (\bar{\bf v}\times\hat{\bf 
B})-\bar{\bf j}_B
\eq

\paragraph{Linear gravity terms}
The gravity induced current densities used in this equation and in 
(\ref{eomei2}) are to first order:
\bqa
{\bf j}_E &=&
- \half \hat{B}_x \frac{\partial}{\partial z}
\left(\begin{array}{c}
0 \\
\bar{h} \\
0
\end{array}
\right) \\\nonumber
{\bf j}_B &=& - \half \hat{B}_x \frac{\partial}{\partial t}
\left(\begin{array}{c}
\bar{h} \\
0 \\
0
\end{array}
\right) 
\eqa

\newpage
\subsection{{\sc mhd} wave dispersion relation}\label{dispersion} 

Finally, it is possible to derive a dispersion relation for the 
plasma waves, generated by the {\sc gw}s that perturb the plasma. 
To obtain such an equation for the combined matter plus 
electrodynamic energy density content, the Faraday tensor no longer 
suffices, as it did in Sec. \ref{gwemw}. What is needed now is 
the second covariant derivative of the total stress-energy tensor, which 
amounts to substituting (minus) the time derivative of the energy conservation 
equation (\ref{energy}) into the divergence of the equations of 
motion (\ref{eomei2}). This is somewhat similar to the procedure used 
when deriving magnetic wave equations by substituting the time 
derivative of Faraday's law into the curl of Amp\`eres law (or the other 
way around to obtain electric waves) .

To first order, the wave equation becomes:

\bqa\label{nablnabl}
\nabla_a \nabla_b T^{ab} &=& 
-\afg{} \haak{\nabla_b T^{0b}} + 
\nabla \cdot \haak{\nabla_b T^{\alpha b}} \\\nonumber
\afgg{\bar{\mu}} - \nabla^2 \bar{p} &=&  
- \frac{1}{4\pi} \nabla\cdot\haak{\nabla\times\bar{\bf 
B}}\times\hat{\bf B} + 
\frac{1}{4\pi}\nabla\cdot(\bar{\bf \mbox{\j}}_E\times\hat{\bf B}) 
\\\nonumber
&=& \frac{\hat{B}_x}{4\pi}\nabla^2 \bar{\bf B}  
- \frac{\hat{B}_x^2}{8\pi} \nabla^2 \bar{\bf h}
\eqa
or, with the definition of the sound velocity:
\bq\label{oef1}
\krul{\afgg{} - c_s^2\nabla^2}\bar{\mu}  =
\frac{\hat{B}_x}{4\pi}\nabla^2 \bar{\bf B} -  \frac{\hat{B}_x^2}{8\pi} \nabla^2 \bar{\bf h}  
\eq
To eliminate $\bar{\bf B}$, differentiate (\ref{max}) with respect to time and insert 
$\nabla\cdot\bar{\bf v}$ from the equation of mass conservation.

\bqa\label{oef2}
\afgg{\bar{\bf B}} &=& 
-\hat{\bf B}\afg{}\nabla\cdot\bar{\bf v} - \afg{\bar{\bf 
\mbox{\j}}_B} \\\nonumber
&=& \hat{\bf B}\afgg{}\krul{\frac{\bar{\mu}}{x} + \frac{\bar{h}}{2}}
\eqa
Before continuing, first define the relativistic
Alfv\'{e}n speed by: 
\bq
\frac{1}{u_A^2} = \frac{1}{c^2} + \haak{\frac{\hat{B}_x^2}{4\pi 
x}}^{-1} \mdoosje{or}
\frac{\hat{B}_x^2}{4\pi x} = \frac{u^2_A}{1-u^2_A/c^2} \mdoosje{with} x=\mu + p
\eq
This expression reduces to $\frac{\hat{B}_x^2}{4\pi x} = u^2_A$ 
as long as we are discussing the {\sc mhd} approximation, where the displacement current is negligible with respect to the current density. One has to keep in mind though, that in the limit of a very thin plasma ($\downarrow $ vacuum) with a strong magnetic field, the asymptotic behaviour of $u_A^2$ is that it goes to $c$ and of course not to infinite speed. 

To arrive at the final dispersion relation for plasma waves excited by a incident gravitational wave, allow for damping, so the energy of the {\sc gw} can dissipate into the plasma. In other words, keep the frequency and wavenumber of the plasma waves unrestricted and try perturbations of the form: 
\bqa\label{perturb}
h &=& \bar{h}\mbox{e}^{i(k_g z-\omega_g t)} \\\nonumber
\mu_{\mbox{\tiny tot}} &=& \mu + \bar{\mu}\mbox{e}^{i(kz-\omega t)} 
\\\nonumber
{\bf B}_{\mbox{\tiny tot}} &=& \hat{\bf B} + \bar{\bf 
B}\mbox{e}^{i(k z-\omega t)}
\eqa
Inserting this into (\ref{oef1}) and (\ref{oef2}), results in the dispersion relations:
\bq\label{oef3}
\{-\omega^2 + k^2(c_s^2 + u_A^2) \} \bar{\mu} =
\frac{\hat{B}_x^2 \omega_g^2}{8\pi} \krul{\frac{k_g^2}{\omega_g^2} - \frac{k^2}{\omega^2}}\bar{h}
\eq
A more physical form of this equation is found after the left-hand-side of the equation is made dimensionless 
(just as the right-hand-side in $\bar{h}$) by normalizing the energy density perturbation to the total energy density: $\bar{\epsilon}=\bar{\mu}/x$. 

Dividing both sides of (\ref{oef3}) by $x$, leads to the final dispersion relation:

\begin{center}
\fbox{\hspace{0.3cm}$
\{-\omega^2 + k^2(c_s^2 + u_A^2) \} \bar{\epsilon} =
\half u_A^2 \omega_g^2 \krul{\frac{k_g^2}{\omega_g^2} - \frac{k^2}{\omega^2}}\bar{h}
$\hspace{0.3cm}}
\end{center}
The characteristics of this equation are discussed in the next subsection.

\subsection{Summary}
In this section, the equations of total energy density conservation and motion were derived in a tedrad system describing the influence of a passing gravitational wave on the underlying metric. It was found that these equations, to first order in $\bar{h}$, have exactly the same form as in a flat metric. 

The coupling with the {\sc gw} comes in through Maxwell's equations, used to eliminate ${\bf B}$, with extra gravity induced terms. Standard {\sc mhd} approximations of infinite conductivity and negligible displacement currents were used to eliminate ${\bf E}$ from the equations. Finally, a wave equation was derived from the second covariant derivative of the total energy-momentum tensor for a perfect fluid in a electromagnetic field.

The dispersion equation that resulted from this, satisfies the expected limiting behaviour:
\begin{itemize}
    \item[$\blacklozenge $] In the limit of vanishing {\sc gw}s, $\bar{h}\rightarrow 0$, it reduces to a common fast magneto\-acoustic 
    plasma wave with $\omega^2 = k^2(c_s^2 + u_A^2)$.
    \item[$\blacklozenge $] Without damping (when $k=k_{g}$ and 
    $\omega=\omega_{g}$), the interaction disappears, and one finds the same magnetosonic solution.
    \item[$\blacklozenge $] In the limit of vanishing density, $\bar{\epsilon} \rightarrow 0$, the displacement current is no longer neglegible, and one has to use the relativistic expressions for $u_A$ and $c_s$, that go to $c$ and $0$ respectively in the vacuum limit. The `plasma' dispersion relation therefore tends to $k=\omega$ (vacuum solution) and one retrieves the dispersion relation for a gravitational wave in vacuum, $k_g=\omega_g$.
    \item[$\blacklozenge $] Finally, the strength of the coupling depends on the square of the magnetic background field through the Alfv\'en speed and on the square of the frequency of the driving {\sc gw}s (compare to the linear dependence on the frequency of the amplitudes of the {\sc emw}s generated by {\sc gw}s). 
\end{itemize}

For a class of {\em gamma ray bursts}, powered by merging neutron star binaries, it would be very interesting if even a small fraction of the enormous amounts of energy released in gravitational waves could be dissipated into the surrounding plasma leading to the observed fireball af a {\sc grb}. This might be an alternative to the explanation that these fireballs are fueled purely by the neutrino flux from the merger, which has the complication that the continuous neutrino flux is already much larger than can be explained by the present models. 

More work on this subject has to be done, though, before one can make any bold statements about the true importance of plasma waves generated by gravitational waves in the surroundings of merging neutron stars, leading to gamma ray burst fireballs. Obviously, the calculations should be extended to spherical symmetries, dipolar magnetic fields and spherically decaying gravitational waves. The {\sc emw}s calculated this way will probably look more like Bessel functions than plane waves.
Still, the results of the first order calculations presented in this thesis seem significant enough to motivate further research.

\newpage
\section{Conclusions}\label{conclusions}
In this thesis, several theoretical coupling effects between gravitational waves 
and electromagnetic waves were investigated. Some of these effects will 
be more significant than others in an astrophysical context.\footnote{All 
of the mentioned couplings are {\em only} relevant in astrophysics.}
The thesis started with an estimate of the general coupling between 
{\sc emw}s and {\sc gw}s in an {\sc em} background field, through the Einstein 
field equations. The result of this, and of the subsequent exact 
calculation, was that the coupling efficiency depends on the square 
of the background field strength and the size of the interaction 
region, with however, and extremely small coupling constant ($\sim 
10^{-50}(\mbox{Ns}^{2}/\mbox{kg m})^{2}$).
These conversions therefore occur either in small regions with very 
strong {\sc em} background fields or in weaker fields extending over 
very large distances.

The astrophysical relevance of this process lies in three areas: it could provide indirect means to detect gravitational waves, it offers a possible explanation for the small fluctuations in the cosmic background radiation and it might prove to fuel the fireballs produced by gamma ray bursts. These results are discussed here in some detail.

\subsection{Radio waves from binary mergers and magnetars}
In the vicinity of merging neutron star binaries or quaking 
supernova remnant neutron stars surrounded by a vacuum, {\sc emw}s 
with maximum amplitudes of $10^{12}$V/m and $50$MV/m 
respectively, could, in principle, be excited. The frequencies of 
these waves are $\sim 5$ kHz and $\sim 10$ kHz respectively, viz long radio 
waves, the latter of which might just be detectable with a space based 
radio array.

In the more realistic situation where the neutron stars
are surrounded by a thin plasma, the light generated by the {\sc gw}s 
obeys a slightly different dispersion relation, with 
$k_{l}^{2}=\omega^{2} - (\Delta \omega)^{2}$ instead of the vacuum 
plane wave dispersion relation $k=\omega$, where $\omega$ is the 
driving frequency of the {\sc gw}. 
In other words, {\sc emw}s 
are still produced by the {\sc gw}s, but there is some dispersion of these 
waves caused by the presence of a plasma. This dispersion shifts the wavenumber 
of the {\sc emw}s by a small amount and also effects the polarization 
by slightly changing the linear polarization relation, 
$\bar{B}_{x} = -\bar{E}_{y}$.

Both of these effects vanish, though, in the strong magnetic 
background fields under consideration. These fields suppress the 
electron mobility in the plasma and thereby also the dispersion caused 
by these electrons, resulting in $(\Delta \omega)^{2}\uparrow 0$. 
What is left are, again, plane polarized radio waves with the same 
amplitudes as before in a vacuum.

The presence of a plasma does determine, however, whether the generated 
radio waves are able to travel over astronomical distances to the earth. 
The frequency of, in particular the {\sc emw}s coming from the 
binary mergers, is of the same order as the the interstellar plasma 
frequency. The radio waves will therefore be absorbed by the plasma, 
unless close to the merger, non-linear plasma effects result in 
socalled {\em photon acceleration}. This effect might lead to higher 
harmonics of the {\sc emw}s with the same energy, that might then 
be able to overcome the interstellar plasma.
Light from magnetars could have high enough frequencies to overcome 
the plasma damping without such an additional effect.

From the most optimistic point of view, merging neutron star 
binaries and supernova remnants in our local galaxy and the Virgo 
cluster might, indirectly, be observable in radio waves generated by 
the large amounts of gravitational energy released in these processes.
On the other hand, the highly idealized calculations (incident plane 
{\sc gw}s etc.) have probably led to exaggerated results. More 
detailed (spherical) calculations are needed to examine what remains in more 
realistic situations.

\subsection{Fluctuations in cosmic background radiation}\label{cbr}
Another interesting phenomenon that might be explained by the conversion of {\sc gw}s to {\sc emw}s 
is the fluctuations in the $2.7$ K microwave background radiation ({\sc grb}).

This radiation results from the decoupling era when the universe 
was approximately $20 \times 10^{4}$ years old. The temperature had 
then dropped to $2700$ K resulting in the recombination of the until 
then, ionized matter. 
Because of this recombination, the diffusing `electron mist' 
evaporated and became transparent to photons, leading to a present day, 
constant flux of thermal photons from this event.

The wavelength of these photons has reddened because of the 
expansion of the universe during their travel to earth and
from this red-shift, one can determine that the size of the universe 
has increased with a factor $1000$ since the decoupling. As the 
present size of the universe is supposed to be $10^{27}$ cm (from the 
redshift of quasars), one finds that it was 
$\sim 10^{24}$ cm when the {\sc cbr} was created. 
  
If, at that time, a primordial magnetic field of only $10^{-2}$ Gauss prevailed 
throughout space, a fraction of up to $10^{-3}$ of the {\sc gw} energy could have been 
converted into {\sc emw} energy. According to popular cosmological models, 
this would be enough to explain the observed relative fluctuations of
$10^{-5}$ in the cosmic background radiation, which would obviously be a 
very interesting result. 
 
\subsection{Gamma ray bursts}
The most promising candidate as a source for socalled {\em gamma ray 
bursts} ({\sc grb}s) are merging neutron star binaries. These events 
can produce the required energy to fuel {\sc grb}s and moreover, they 
satisfy the temporal and size constraints set by the observations.
Also, these mergers can be observed as often as once per day, which is 
comparable to the number of detected {\sc grb}s.

A problem in most {\sc grb} models is that the energy released by 
merging binaries is released primarily in {\sc gw}s and not in {\sc em} 
radiation. The conversion of the energy in {\sc gw}s to {\sc em} 
energy would therefore be very interesting. As was discussed 
extensively in this thesis, the direct conversion of {\sc gw}s to 
light is not very effective, and more importantly, results in long 
radio waves, certainly not in gamma radiation. 

A promising alternative is offered by the excitation of 
magnetohydrodynamic plasma waves, 
such as slow or fast magneto-acoustic waves and Alfv\'en waves. In 
Section \ref{dispersion}, it was derived for the first time that, 
in particular, fast 
magnetosonic waves can indeed be generated through 
the interaction with gravitational waves passing through the plasma. 
The effectiveness of this process is proportional to the square of 
the strong magnetic background field and the radio frequency of the 
perturbing {\sc gw}s. In Appendix \ref{non-lin} it is shown, 
following \cite{brodin0600}, that non-linear {\sc gw} effects can 
cause longitudinal Alfv\'en-like waves.

The magnetosonic waves are generated only when there is a wavelength 
(or, in the static case, a frequency) mismatch between the {\sc gw} 
and the excited plasma waves. When this is the case, energy from 
the {\sc gw} can dissipate into the plasma, which can later on 
emmit the energy in the form of {\sc em} (maybe gamma-) radiation.

Again, the work done in this thesis is not sufficient to fully 
describe what is going on in the vicinity of {\sc grb}s, but the 
result that a fraction of the {\sc gw} energy from these sources could 
be converted into the observed {\sc em} radiation is one  
that deserves further research. Such research would, as a first 
improvement, have to include dipolar magnetic fields and spherical {\sc gw} solutions for the 
merging neutron stars.

\newpage
\addcontentsline{toc}{section}{Acknowledgements}
\section*{Acknowledgements}
First of all, I would like to thank my supervisors Prof. Dr. J. Kuijpers and Prof. Dr. G. 't Hooft for giving me the opportunity 
to get involved in this interesting research project on a subject where theoretical physics and high energy astrophysics meet.

Secondly, I am indebted to Ass. Prof. Dr. Papadopoulos (Thess., Greece) for improving my understanding of relativistic, magnetohydrodynamic plasma waves and their dispersion relation. This really helped in writing the last part of my thesis. 

I would also like to thank some (too many to list here) of my fellow students for many usefull discussions on this subject over the past few months. These discussions were very helpfull in organizing my wandering thoughts on several problems I encountered during my thesis work.

Concluding, I would like to thank Drs. S. van der Feest for lots of moral support, and Prof. Dr. R. Oerhle (Mass., U.S.A.) and Hylke Koers for some final corrections and suggestions.

 \hspace{10.5cm}Utrecht, march 2001.
 \newpage

\appendix


\section{Constants for exact {\sc gw} to {\sc emw} calculation}\label{constants}

\paragraph{Region I}


\bqa
A_{I} &=& 0 \\\nonumber
B_{I} &=& (\mbox{e}^{-2ikL}-1) \left[ \right.\\\nonumber
&+ & \kwart \left\{ \frac{\alpha}{k^2} - B_{z}^{(0)} (\xi_{11} + 
\xi_{22}) - E_y^{(0)} (\xi_{00} - \xi_{22})\right\} \\ \nonumber
&+& \half \left\{E_z^{(0)} \xi_{23} + B_{x}^{(0)} \xi_{30} - 
 B_{z}^{(0)} \xi_{10} + E_x^{(0)} \xi_{12} + B_{y}^{(0)} 
\xi_{23})\right\} \left. \right] \\\nonumber
C_{I} &=& 0 \\\nonumber
D_{I} &=& (\mbox{e}^{-2ikL}-1) \left[ \right. \\\nonumber
&+ & \kwart \left\{ \frac{\beta}{k^2}
+  B_{y}^{(0)} (\xi_{11} + \xi_{33}) -  E_z^{(0)} (\xi_{00} - \xi_{33}) 
\right\} \\\nonumber
 &+& \half \left\{ E_x^{(0)} \xi_{13} + B_{y}^{(0)} \xi_{10} - 
B_{x}^{(0)} \xi_{20} \right\} \left. \right]  
\eqa

\paragraph{Region II}

\bqa
A_{II} &=& \kwart \krul{\frac{\alpha}{k^2} + B_{z}^{(0)} (\xi_{11} + 
\xi_{22}) -  E_y^{(0)} (\xi_{00} - \xi_{22})} \\\nonumber
 &+& \half \krul{\frac{\alpha L}{ik} + E_z^{(0)} \xi_{23} + B_{x}^{(0)} 
\xi_{30} - 
 B_{z}^{(0)} \xi_{10} + E_x^{(0)} \xi_{12} - B_{y}^{(0)} \xi_{23}} 
\\\nonumber
B_{II} &=& - \kwart \krul{\frac{\alpha}{k^2} - B_{z}^{(0)} (\xi_{11} + 
\xi_{22}) -  E_y^{(0)} (\xi_{00} - \xi_{22})} \\\nonumber
 &-& \half (E_z^{(0)} \xi_{23} + B_{x}^{(0)} \xi_{30} - 
 B_{z}^{(0)} \xi_{10} + E_x^{(0)} \xi_{12} + B_{y}^{(0)} \xi_{23} 
\\\nonumber
C_{II} &=&  \kwart \krul{\frac{\beta}{k^2} -  B_{y}^{(0)} (\xi_{11} + 
\xi_{33}) -  E_z^{(0)} (\xi_{00} - \xi_{33})} \\\nonumber
 &+& \half \krul{ \frac{\beta L}{ik} + E_x^{(0)} \xi_{13} + B_{y}^{(0)} 
\xi_{10}  
 - B_{x}^{(0)} \xi_{20}} \\\nonumber
D_{II} &=& - \kwart \krul{ \frac{\beta}{k^2} + B_{y}^{(0)} (\xi_{11} + 
\xi_{33}) -  E_z^{(0)} (\xi_{00} - \xi_{33})} \\\nonumber
 &-& \half \krul{ E_x^{(0)} \xi_{13} + B_{y}^{(0)} \xi_{10}  
 - B_{x}^{(0)} \xi_{20}} 
\eqa

\paragraph{Region III}


\bqa
A_{III} &=& \frac{\alpha L}{2ik} \\\nonumber
B_{III} &=& 0 \\\nonumber
C_{III} &=& \frac{\beta L}{2ik} \\\nonumber
D_{III} &=& 0
\eqa
\newpage
\section{Symmetrizing the Maxwell equations}\label{symmetric}
To prove that 
$(\epsilon^{\delta\alpha\gamma}\ric{\beta}{\delta\alpha} + 
\epsilon^{\beta\delta\gamma}\ric{\alpha}{\delta\alpha})B_{\gamma}  =
\epsilon^{\beta\delta\gamma}\ric{\alpha}{\delta\gamma}B_{\alpha}$ 
write:

\bqa
\epsilon^{\delta\alpha\gamma}\ric{\beta}{\delta\alpha}B_{\gamma} + 
\epsilon^{\beta\delta\gamma}\ric{\alpha}{\delta\alpha}B_{\gamma} -
\epsilon^{\beta\delta\gamma}\ric{\alpha}{\delta\gamma}B_{\alpha} &=& 
\\\nonumber
(\eta^{\sigma\alpha}\epsilon^{\beta\delta\gamma} + 
\eta^{\sigma\beta}\epsilon^{\delta\alpha\gamma} - 
\eta^{\sigma\gamma}\epsilon^{\beta\alpha\delta})\Gamma_{\sigma\delta\alpha}B_{\gamma} 
&=& \\\nonumber
(
\eta^{\sigma\alpha}\epsilon^{\gamma\beta\delta} + 
\eta^{\sigma\beta}\epsilon^{\alpha\gamma\delta} - 
\eta^{\sigma\gamma}\epsilon^{\beta\alpha\delta}
)\Gamma_{\sigma\delta\alpha}B_{\gamma} &=& 0
\eqa
So one has to prove that the expression in brackets is either 
symmetric in $\sigma\delta$ (since $\Gamma_{\sigma\delta\alpha}$ is 
skewsymmetric in these indices) or vanishes. 

Distinguish between two cases:
\paragraph{Case 1:} For $\alpha \neq \beta \neq \gamma$ the first term gives 
$\sigma = \delta$ and the other terms vanish,
\paragraph{Case 2:} For $\alpha = \gamma \neq \beta$ the second term is zero and 
the other terms cancel.

\paragraph{Conclusion}:
\bq
(\epsilon^{\delta\alpha\gamma}\ric{\beta}{\delta\alpha} + 
\epsilon^{\beta\delta\gamma}\ric{\alpha}{\delta\alpha})B_{\gamma} = 
\epsilon^{\beta\delta\gamma}\ric{\alpha}{\delta\gamma}B_{\alpha}
\eq
which is nice, because it makes the inhomogeneous Maxwell equations 
symmetric with respect to the homogeneous ones.

\newpage
\section{Non-linear effects}\label{non-lin}

It is to be expected that close to a source that emits strong {\sc 
gw}'s such as magnetars and neutron star binaries, non linear {\sc 
gw} effects might become important. Consider, following 
\cite{brodin0600}, a plane fronted parallel, linearly polarized {\sc 
gw}:


\bq\label{nlgw}
ds^2 = -dt^2 + a(z-t)^2 dx^2 + b(z-t)^2 dy^2 + dz^2
\eq

with $z-t=u$ and $ab_{uu} + a_{uu}b =0$. Using the same procedure as 
before, introduce an {\sc onf}, which in this case is a canonical 
Lorentz frame:


\bqa
e_{(0)}^{\quad i} &=& (1, 0, 0, 0)) \\\nonumber
e_{(1)}^{\quad i} &=& (0, a^{-1}, 0, 0) \\\nonumber
e_{(2)}^{\quad i} &=& (0, 0, b^{-1}, 0) \\\nonumber
e_{(3)}^{\quad i} &=& (0, 0, 0, 1) 
\eqa

In this metric the gravity induced charge densities do not 
vanish, and the induced current densities have longitudinal 
components:


\bqa
\rho_E &=& -(\ln ab)_u E^3 \\\nonumber
\rho_B &=& -(\ln ab)_u B^3 \\\nonumber
{\bf j}_E &=& -(\ln b)_u (E^1 - B^2)\ted_1 - (\ln a)_u (E^2 + 
B^1)\ted_2 - (\ln ab)_u E^3\ted_3\\\nonumber
{\bf j}_B &=& -(\ln b)_u (E^2 + B^1)\ted_1 + (\ln a)_u (E^1 - 
B^2)\ted_2 - (\ln ab)_u B^3\ted_3\\\nonumber
\afg{n} + \nabla \cdot (n{\bf v}) &=& -n(\ln ab)_u (1- v_3 ) 
\\\label{eqmotion3}
\haak{\afg{} + {\bf v}\cdot \nabla }{\bf v} &=& \frac{q}{m} ({\bf E} 
+ {\bf v}\times {\bf B}) \\\nonumber
&+& ((\ln a)_u v_1 \ted_1 + (\ln b)_u v_2 \ted_2)(1-v_3) + ((\ln a)_u 
v_1^2 + (\ln b)_u v_2^2)\ted_3
\eqa

If we assume that the {\sc gw}s are approximately periodic and still 
have small amplitudes, the logarithmic factors that appear in these 
equations can be approximated by $a(u) = 
\sum_{n=-\infty}^{\infty}\hat{a}_n \exp{in\omega u}$ and $b(u) = 
\sum_{n=-\infty}^{\infty}\hat{b}_n \exp{in\omega u}$. To second order 
in $\bar{h}$ we find (see \cite{brodin0600} for details):


\bqa
(\ln a)_u &=& i\omega \bar{h} \mbox{e}^{i\omega (z-t)} - \half 
i\omega \bar{h}^2 \mbox{e}^{2 i\omega (z-t)} + \mbox{c.c.}\\\nonumber
(\ln b)_u &=& -i\omega \bar{h} \mbox{e}^{i\omega (z-t)} - \half 
i\omega \bar{h}^2 \mbox{e}^{2 i\omega (z-t)}+ \mbox{c.c.}\\\nonumber
(\ln ab)_u &=& - i\omega \bar{h}^2 \mbox{e}^{2 i\omega (z-t)} - 
\frac{1}{8} i\omega \bar{h}^4 \mbox{e}^{4 i\omega (z-t)}+ 
\mbox{c.c.}\\\nonumber
\eqa 

For a {\sc gw} moving to an initially unperturbed plasma with ${\bf 
E}_0 = {\bf B}_0 = \partial n_0/\partial t =0$, (\ref{maxje}) and 
(\ref{eqmotion3}), to first order, reduce to:


\bqa
\afg{E_z} &=& (\ln ab)_u E_z - 4\pi q n_0 v_z \\\nonumber
\afg{v_z} &=& \frac{q}{m} E_z \mdoos{and} \\\nonumber
\haak{\afgg{} + \omega_p^2}E &=&\afg{}(\ln ab)_u E 
\eqa

Obviously, the flat space plasma oscillations are changed by the {\sc 
gw}, especially at the resonance frequency $\omega_p = \omega$, where 
the parametric excitation of longitudinal plasma oscillations is 
given by:


\bq
\ddt{\bar{E}} = -\half i \omega \bar{h}^2\mbox{e}^{2i\omega z}  
\bar{E}^{\ast}
\eq

with a growth rate of:

\bq
\Gamma = \half \omega |\bar{h}^2 |
\eq

\newpage
\nocite{*}
\bibliographystyle{plain}
\bibliography{mybiblio}

\end{document}